\documentclass{mrv9x6}

\usepackage{makeidx,wsminitoc}

\makeindex

\makeatletter  %
\nomtcrule
\def\mtc@bottom@rule{\vskip 5pt \hrule height \z@ width \columnwidth \kern2.6\p@}
\def\mtifont{\small\bf}           
\renewcommand{\mtcfont}{\small\rm} 
\renewcommand{\mtcSfont}{\small\rm}

\renewcommand*\l@section{\@dottedtocline{1}{1em}{1.3em}}
\renewcommand*\l@subsection{\@dottedtocline{2}{2.3em}{2em}}
\renewcommand*\l@subsubsection{\@dottedtocline{3}{4.3em}{2.6em}} 
\makeatother

\begin{document}

\dominitoc       

\makeatletter
\renewcommand\thedefinition{\thesection.\arabic{definition}}
\newtheorem{defn}{Definition}[section]
\newtheorem{theo}{Theorem}[section]
\newtheorem{lem}{Lemma}[section]
\newtheorem{exam}{Example}[section]
\newtheorem{cor}{Corollary}[section]
\newtheorem{prop}{Proposition}[section]
\newtheorem{rem}{Remark}[section]
\def\x{{\bf x}}
\def\y{{\bf y}}
\newcommand{\X}{{\cal X}}
\newcommand{\F}{\mbox{\bf F}}
\newcommand{\R}{\mbox{\bf R}}
\def\L{{\cal L}}
\def\FX{{\F_q(\X)}}
\def\wt{{\rm wt}}
\def\nP{\nu_P}
\makeatother


\setcounter{page}{5}



\setcounter{page}{1}


\chapter[INTRODUCTION TO MARKOV CHAIN MONTE CARLO SIMULATIONS AND THEIR
STATISTICAL ANALYSIS]{INTRODUCTION TO MARKOV CHAIN MONTE CARLO SIMULATIONS 
AND THEIR STATISTICAL ANALYSIS}

\markboth{B.A. Berg}{Markov Chain Monte Carlo Simulations
and their Statistical Analysis}

\author{Bernd A. Berg}

\address{Department of Physics\\
Florida State University\\
Tallahassee, Florida 32306-4350, USA\\
and\\
School of Computational Science\\
Florida State University\\
Tallahassee, Florida 32306-4120, USA}

\begin{abstract}
This article is a tutorial on Markov chain Monte Carlo simulations 
and their statistical analysis. The theoretical concepts are 
illustrated through many numerical assignments from the author's
book \cite{Berg} on the subject. Computer code (in Fortran) is 
available for all subjects covered and can be downloaded from the web. 
\end{abstract}

\vspace*{24pt}


\minitoc         

\section{Introduction}

Markov chain Monte Carlo (MC) simulations started in earnest with 
the 1953 article by Nicholas Metropolis, Arianna Rosenbluth, Marshall
Rosenbluth, Augusta Teller and Edward Teller~\cite{Me53}. Since
then MC simulations have become an indispensable tool with applications
in many branches of science. Some of those are reviewed in the proceedings
\cite{Gu04} of the 2003 Los Alamos conference, which celebrated the 50th 
birthday of Metropolis simulations. 

The purpose of this tutorial is to provide an overview of basic 
concepts, which are prerequisites for an understanding of the more 
advanced lectures of this volume. In particular the lectures by
Prof. Landau are closely related.

The theory behind MC simulations is based on statistics and the
analysis of MC generated data is applied statistics. Therefore, 
statistical concepts are reviewed first in this 
tutorial. Nowadays abundance of computational power implies also
a paradigm shift with respect to statistics: Computationally 
intensive, but conceptually simple, methods belong at the 
forefront. MC simulations are not only relevant for simulating
models of interest, but they constitute also a valuable tool for
approaching statistics.

The point of departure for treating Markov chain MC simulations is
the Metropolis algorithm for simulating the Gibbs canonical ensemble.
The heat bath algorithm follows. To illustrate these methods our 
systems of choice are discrete Potts and continuous $O(n)$ models.
Both classes of models are programmed for arbitrary dimensions
($d=1,2,3,4,\dots$). On the advanced side we introduce multicanonical
simulations, which cover an entire temperature range in a single
simulation, and allow for direct calculations of the entropy and
free energy. 

In summary, we consider Statistics, Markov Chain Monte Carlo 
simulations, the Statistical Analysis of Markov chain data and, 
finally, Multicanonical Sampling. This tutorial is abstracted from 
the author's book on the subject~\cite{Berg}. Many details, which 
are inevitably ommitted here, can be found there.

\section{Probability Distributions and Sampling}

A \index{sample space}{\bf sample space} is a set of points
or elements, in natural sciences called {\bf measurements} or 
{\bf observations}, whose occurrence depends on chance.
Carrying out independent repetitions of the same experiment is called 
\index{sampling}{\bf sampling}. The outcome of each experiment provides 
an event called data point. In $N$ such experiments we may find the 
event $A$ to occur with {\bf frequency} $n$, $0\le n\le N$. The 
\index{probability} {\bf probability} assigned to the event $A$ 
is a number $P(A)$, $0\le P(A) \le 1$, so that
\begin{equation} \label{fdp}
P(A)\ =\ \lim_{N\to\infty} {n\over N} .
\end{equation}
This equation is sometimes called the {\bf frequency definition 
of probability}. 

Let us denote by $P(a,b)$ the probability that $x^r\in [a,b]$ where 
$x^r$ is a continuous \index{random variable}{\bf random variable} 
drawn in the interval $(-\infty,+\infty)$ with the {\bf probability 
density} $f(x)$. Then,
\begin{equation} \label{Pab}
P(a,b) = \int_a^b f(x)\ dx .
\end{equation}
Knowledge of all probabilities $P(a,b)$ implies 
\begin{equation} \label{pd}
f(x) = \lim_{y\to x^-} {P(y,x) \over x-y}\ \ge\ 0\ .
\end{equation}
The {\bf (cumulative) distribution function} of the random variable 
$x^r$ is defined as 
\begin{equation} \label{df}
F(x) = P(x^r\le x) = \int_{-\infty}^x f(x)\,dx\,.
\end{equation}
A particularly important case is the {\bf uniform probability
distribution} for random numbers between $[0,1)$, 
\begin{equation} \label{upd}
u(x) = \begin{cases} 1~~{\rm for}~~0\le x < 1;\\
                     0 ~~{\rm elsewhere}.  
       \end{cases}
\end{equation}
Remarkably, the uniform distribution allows for the construction of 
general probability distributions.  Let
$$ y = F(x) = \int_{-\infty}^x f(x')\,dx' $$
and assume that the inverse $x=F^{-1}(y)$ exists. For $y^r$ being 
a uniformly distributed random variable in the range $[0,1)$ it
follows that
\begin{equation} \label{gd}
x^r = F^{-1} (y^r)
\end{equation}
is distributed according to the probability density $f(x)$. 

The \index{Gaussian distribution}{\bf Gaussian} or 
\index{normal distribution}{\bf normal distribution} is of major 
importance. Its probability density is 
\begin{equation} \label{gpd}
 g(x) = {1\over \sigma \sqrt{2\pi}}\, e^{-x^2/(2\sigma^2)}
\end{equation}
where $\sigma^2$ is the \index{variance}{\bf variance} and $\sigma > 0$ 
the \index{standard deviation}{\bf standard deviation}. The Gaussian 
distribution function $G(x)$ is related to that of variance 
$\sigma^2=1$ by 
\begin{equation} \label{gdf}
G(x) = \int_{-\infty}^x g(x')\,dx' = {1\over \sqrt{2\pi}}
\int_{-\infty}^{x/\sigma} e^{-(x'')^2/2}\, dx'' = {1\over 2} +
{1\over 2}\, {\rm erf} \left( {x\over \sigma\sqrt{2}} \right)\ .
\end{equation}
In principle we could now generate {\bf Gaussian random numbers}
according to Eq.~(\ref{gd}). 
However, the numerical calculation of the inverse error function 
is slow and makes this an impractical procedure. Much faster is 
to express the product probability density of two independent 
Gaussian distributions in polar coordinates
$${1\over 2\pi\,\sigma^2}\,e^{-x^2/(2\sigma^2)}\,e^{-y^2/(2\sigma^2)}\,
dx\,dy ={1\over 2\pi\,\sigma^2}\,e^{-r^2/(2\sigma^2)}\,d\phi\,rdr\, ,$$
and to use the relations
\begin{equation} \label{grv}
x^r = r^r\, \cos \phi^r ~~{\rm and}~~ y^r = r^r\, \sin \phi^r\ .
\end{equation}

\section{Random Numbers and Fortran Code}

According to Marsaglia and collaborators \cite{MaZa90} a list of 
desirable properties for (pseudo) random number generators is:

\begin{description}
\item{(i)} {\it Randomness}. The generator should pass stringent tests
for randomness.

\item{(ii)} {\it Long period}. 

\item{(iii)} {\it Computational efficiency}. 

\item{(iv)} {\it Repeatability}. Initial conditions (seed values) 
completely determine the resulting sequence of random variables.

\item{(v)} {\it Portability}. Identical sequences of random variables may 
be produced on a wide variety of computers (for given seed values).

\item{(vi)} {\it Homogeneity}. All subsets of bits of the numbers are
random.
\end{description}

Physicists have added a number of their applications as new tests
(e.g., see \cite{VaAl95} and references therein). In our program package
a version of the random number generator of Marsaglia and collaborators 
\cite{MaZa90} is provided. Our corresponding Fortran code consists of 
three subroutines: 

\begin{description}
\item {\tt rmaset.f} to set the initial state of the random number 
      generator.
\item {\tt ranmar.f} which provides one random number per call.
\item {\tt rmasave.f} to save the final state of the generator. 
\end{description}

In addition, {\tt rmafun.f} is a Fortran function version of 
{\tt ranmar.f} and calls to these two routines are freely 
interchangeable. Related is also the subroutine {\tt rmagau.f}, 
which generates two Gaussian random numbers. 

The subroutine {\tt rmaset.f} initializes the generator to mutually
independent sequences of random numbers for distinct pairs of 
\begin{equation} \label{seeds}
-1801 \le {\tt iseed1} \le 29527 ~~{\rm and}~~
-9373 \le {\tt iseed2} \le 20708 \ .
\end{equation}
This property makes the generator quite useful for parallel processing. 

\subsection{How to get and run the Fortan code}

\begin{figure}[tb]
\vspace{5pc}
 \centerline{\hbox{ \psfig{figure=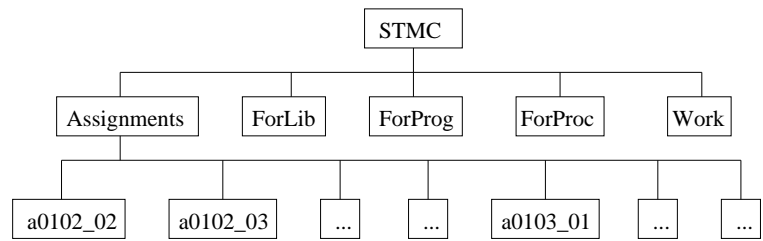,width=10cm} }}
 \caption{The Fortran routines are provided and prepared to run in 
  the tree structure of folders depicted in this figure. This 
  tree unfolds from the downloaded file. } 
 \label{fig_fort}
\end{figure}

\noindent To {\bf download} the Fortran code book visit the website
$$ {\tt http://b\_ berg.home.comcast.net/} $$
and follow the instructions given there. If the above link should be
unavailable, visit the author's homepage which is presently located at
\smallskip

\centerline{ {\tt http://www.hep.fsu.edu/\~\ \!\!berg}\ .}
\smallskip

After installation the directory tree shown in Fig.~\ref{fig_fort} is
obtained. {\tt ForLib} contains a library of functions and 
subroutines which is closed in the sense that no reference to 
non-standard functions or subroutines outside the library is ever 
made. Fortran programs are contained in the folder {\tt ForProg} and 
procedures for interactive use in {\tt ForProc}. It is {\bf recommended} 
to leave the hyperstructure of program dependencies introduced between 
the levels of the STMC directory tree intact. Otherwise, complications 
may result which require advanced Fortran skills.

{\bf Assignment: Marsaglia random numbers.} Run the program {\tt mar.f} 
to reproduce the following results: 
\begin{verbatim}
  RANMAR INITIALIZED.               MARSAGLIA CONTINUATION.
  idat, xr = 1  0.116391063         idat, xr = 1  0.495856345
  idat, xr = 2  0.96484679          idat, xr = 2  0.577386141
  idat, xr = 3  0.882970393         idat, xr = 3  0.942340136
  idat, xr = 4  0.420486867         idat, xr = 4  0.243162394
  extra xr =    0.495856345         extra xr =    0.550126791
\end{verbatim}
Understand how to 
re-start the random number generator and how to perform different 
starts when the continuation data file {\tt ranmar.d} does not exist. 
You find {\tt mar.f} in {\tt ForProg/Marsaglia} and it includes 
subroutines from {\tt ForLib}. To compile properly, {\tt mar.f} has to 
be located {two levels down from a root directory {\tt STMC}}. The
solution is given in the folder {\tt Assignments/a0102\_02}.

\section{Confidence Intervals and Heapsort}

Let a distribution function $F(x)$ and $q$, $0\le q\le 1$ be given.
One defines {\bf ${\bf q}$-tiles} (also called {\bf quantiles} 
or {\bf fractiles}) $x_q$ by means of
\begin{equation} \label{xq}
 F(x_q)\ =\ q\ .
\end{equation}
The {\bf median} $x_{1\over 2}$ is often (certainly not always) the 
{\bf typical} value of the random variable $x^r$. 

\begin{figure}[tb]
\vspace{5pc}
 \centerline{\hbox{ \psfig{figure=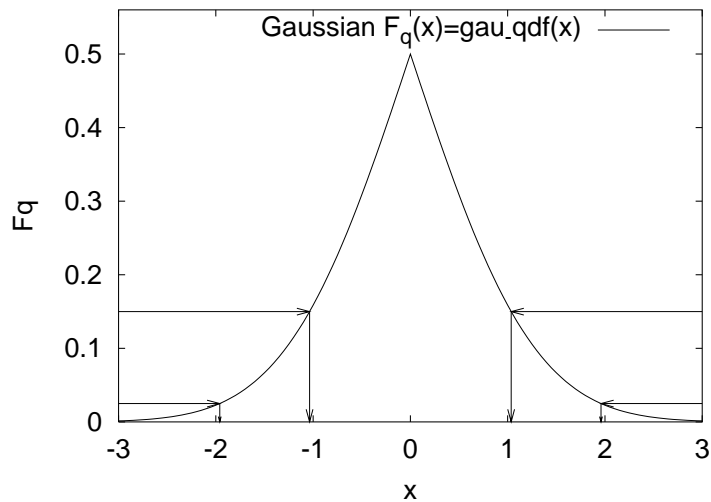,width=10cm} }}
\caption{\small Gaussian peaked distribution function
and estimates of $x_q$ for the 70\% (approximately $1\,\sigma$) 
and 95\% (approximately $2\,\sigma$) confidence intervals.}
 \label{fig_gqdf}
\end{figure}

Example: For the normal distribution the precise probability content 
of the confidence intervals
$$ [x_q,x_{1-q}] = [-n\sigma,n\sigma]\ \ {\rm for}\ \ n=1,2 $$
is $p=1-2q=68.27$\% for one $\sigma$ and $p=1-2q=95.45$\% for 
two $\sigma$. 

The {\bf peaked distribution function} 
\begin{equation} \label{qdf}
F_q(x) = \begin{cases} 
         F(x)\ {\rm for}\ F(x) \le {1\over 2},\\
         1 - F(x)\ {\rm for}\ F(x) > {1\over 2}.
         \end{cases} 
\end{equation}
provides a useful way to visualize probability intervals of a 
distribution. It is illustrated in Fig.~\ref{fig_gqdf} for the
Gaussian distribution.

Sampling provides us with an empirical distribution function and in 
practice the problem is to estimate confidence intervals from the 
empirical data. Assume we generate $n$ random numbers $x_1,...,x_n$ 
independently
according to a probability distribution $F(x)$. The $n$ random numbers 
constitute a {\bf sample}.  We may re-arrange the $x_i$ in increasing 
order. Denoting the smallest value by $x_{\pi_1}$, the next smallest 
by $x_{\pi_2}$, etc., we arrive at
\begin{equation} \label{order_stat}
 x_{\pi_1} \le x_{\pi_2} \le \dots \le x_{\pi_n} 
\end{equation}
where $\pi_1,\dots , \pi_n$ is a permutation of $1,\dots ,n$. Each 
of the $x_{\pi_i}$ is called an {\bf order statistic}.
An estimator for the distribution function $F(x)$ is the
{\bf empirical distribution function}
\begin{equation} \label{edf}
\overline{F} (x) = {i\over n}~~~{\rm for}~~~x_{\pi_i} \le x <
x_{\pi_{i+1}},\ i=0, 1,\dots , n-1, n
\end{equation}
with the definitions $x_{\pi_0}=-\infty$ and $x_{\pi_{n+1}}=+\infty$.

To calculate $\overline{F}(x)$ and the corresponding peaked 
distribution function, one needs an efficient way to {\bf sort} 
$n$ data values in ascending (or descending) order. This is
provided by the {\bf heapsort}, which relies on two steps: First 
the data are arranged in a heap, then the heap is sorted.
A {\bf heap} is a partial ordering so that the number at the top is 
larger or equal than the two numbers in the second row, provided at 
least three numbers $x_i$ exist. More details are given in \cite{Berg}.
The computer time needed to succeed with this sorting process grows 
only like $n\,\log_2 n$, because there are $\log_2 n$ levels in the 
heap, see Knuth~\cite{Kn68} for an exhaustive discussion of sorting 
algorithms. 

\section{The Central Limit Theorem and Binning }

How is the sum of two independent random variables
\begin{equation} \label{two_r}
 y^r\ =\ x_1^r + x_2^r\ . 
\end{equation}
distributed? We denote their probability density of $y^r$ by $g(y)$. 
The corresponding cumulative distribution function is given by 
$$ G(y) = \int_{x_1+x_2 \le y} f_1(x_1)\ f_2(x_2)\ dx_1\ dx_2 
        = \int_{-\infty}^{+\infty} f_1(x)\ F_2(y-x)\ dx $$
where $F_2(x)$ is the distribution function of the random variable 
$x_2^r$. We take the derivative and obtain the probability 
density of $y^r$
\begin{equation} \label{two_pd}
 g(y)\ =\ {dG(y)\over dy}\  
       =\ \int_{-\infty}^{+\infty} f_1(x)\ f_2(y-x)\ dx\ . 
\end{equation}
The probability density of a sum of two independent random variables 
is the {\bf convolution of the probability densities} of these random 
variables.  

Example: Sums of uniform random numbers, corresponding to the sums 
of an uniformly distributed random variable $x^r\in (0,1]$:

(a) Let $y^r=x^r+x^r$, then
\begin{equation} \label{two_upd}
 g_2(y)\ =\ \begin{cases} y ~~~~~{\rm for}~~~ 0\le y\le 1, \\
                    2-y  ~~{\rm for}~~~ 1\le y\le 2, \\
                    0   ~~~{\rm elsewhere} .
            \end{cases}
\end{equation}

(b) Let $y^{\,r}=x^r+x^r+x^r$, then
\begin{equation} \label{three_upd}
 g_3(y)\ =\ \begin{cases} y^2/2 ~~~~~~{\rm for}~~ 0\le y\le 1, \\
              (-2y^2+6y-3)/2 ~~{\rm for}~~ 1\le y\le 2, \\
                 (y-3)^2/2 ~~~~{\rm for}~~ 2\le y\le 3, \\
                  0   ~~~{\rm elsewhere} . 
            \end{cases}
\end{equation}

\begin{figure}[tb]
\vspace{5pc}
 \centerline{\hbox{ \psfig{figure=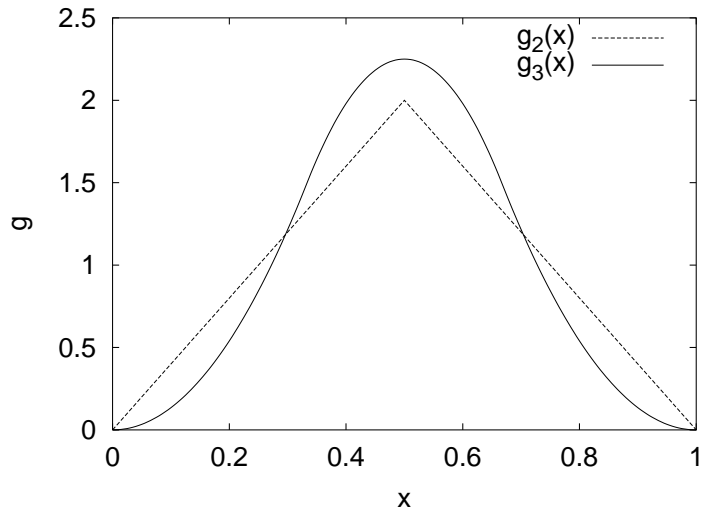,width=10cm} }}
\caption{Probability densities for the arithmetic means 
 of two and three uniformly distributed random variables, 
 $\widehat{g}_2(\overline{x})$ and $\widehat{g}_3(\overline{x})$, 
 respectively.} \label{fig_mean_u}
\end{figure}

The convolution~(\ref{two_pd}) takes on a simple form in 
{\bf Fourier space}. In statistics the {\bf Fourier transformation} 
of the probability density is known as {\bf characteristic function},
defined as the expectation value of $e^{itx^r}$:
\begin{equation} \label{chfct}
 \phi(t)\ =\ \langle e^{itx^r} \rangle\ =\ 
 \int_{-\infty}^{+\infty} e^{itx}\, f(x)\, dx\ .
\end{equation}
A straightforward calculation gives
\begin{equation} \label{chfct_gau}
 \phi(t)\ =\ \exp\left[ -{1\over 2}\,{\sigma_x^2\over N}\,t^2\right]
\end{equation}
for the characteristic function of the Gaussian probability 
density~(\ref{gpd}). The characteristic function is particularly useful 
for investigating sums of random variables, $y^r = x_1^r + x_2^r$:
\begin{eqnarray} \label{two_cf}
 \phi_y(t) &=& \langle e^{(itx_1^r+itx_2^r)} \rangle \\ \nonumber
 &=&  \int_{-\infty}^{+\infty} \int_{-\infty}^{+\infty} 
  e^{itx_1}\ e^{itx_2}\ f_1(x_1)\ f_2(x_2)\ dx_1\ dx_2
  = \phi_{x_1}(t)\ \phi_{x_2}(t)\ .
\end{eqnarray}
{\bf The characteristic function of a sum of random variables is the 
product of their characteristic functions.} The result generalizes
immediately to $N$ random variables
$ y^r =\x_1^r + \dots + x_N^r$. 
The characteristic function of $y^r$ is
\begin{equation} \label{N_cf}
 \phi_y(t)\ =\ \prod_{i=1}^N \phi_{x_i}(t)
\end{equation}
and the probability density of $y^r$ is the Fourier back-transformation
of this characteristic function
\begin{equation} \label{sum_pd}
g(y)\ =\ {1\over 2\pi} \int_{-\infty}^{+\infty} dt\, e^{-ity}\,
\phi_y(t)\ .
\end{equation}
The {\bf probability densitiy of the sample mean} is obtained as follows:
The arithmetic mean of $y^r$ is $\overline{x}^{\, r}=y^r/N$.
We denote the probability density of $y^r$ by $g_N(y)$ and the probability 
density of the arithmetic mean by $\widehat{g}_N(\overline{x})$. They are 
related by 
\begin{equation} \label{mean_N_pd}
 \widehat{g}_N (\overline{x})\ =\ N\, g_N(N\overline{x})\ . 
\end{equation}
This follows by substituting $y=N\overline{x}$ into $g_N(y)\,dy$:
$$ 1 = \int_{-\infty}^{+\infty} g_N(y)\,dy = 
\int_{-\infty}^{+\infty} g_N(N\overline{x})\,2d\overline{x} = 
\int_{-\infty}^{+\infty} \widehat{g}_N(\overline{x})\,d\overline{x}\ .$$
Fig.~\ref{fig_mean_u} illustrates equation (\ref{mean_N_pd}) for the
sums of two (\ref{two_upd}) and three (\ref{three_upd}) uniformly 
distributed random variables.
This suggests that sampling leads to convergence of the mean
by reducing its variance. We use the characteristic function
$\phi_y(t)\ =\ \left[ \phi_x(t) \right]^N$
to understand the general behavior. The characteristic function 
for the corresponding arithmetic average is
$$ \phi_{{\overline x}}(t)\ =\ \int_{-\infty}^{+\infty} 
   d\overline{x}\, e^{it\overline{x}}\, \widehat{g}_N(\overline{x})\ 
=\ \int_{-\infty}^{+\infty} dy\, \exp \left( i\,{t\over N}\, y\right)\,
    g_N (y)\ . $$
Hence,
\begin{equation} \label{mean_chf}
 \phi_{{\overline x}}(t)\ =\ \phi_y \left( {t\over N} \right)\
 =\ \left[ \phi_x \left( {t\over N} \right) \right]^N\ .
\end{equation}
To simplify the equations we restrict ourselves to ${\widehat x}=0$. 
Let us consider a probability density $f(x)$ and assume that its 
moment exists, implying that the characteristic function is a least 
two times differentiable, so that
\begin{equation} \label{approx_chf}
 \phi_x (t)\ =\ 1\ -\ {\sigma_x^2 \over 2}\ t^2\ +\ {\cal O} (t^3)\ .
\end{equation}
The leading term reflects the the normalization of the probability density
and the first moment is $\phi' (0)=\widehat{x}=0$. The characteristic 
function of the mean becomes 
$$\phi_{\overline x} (t)\ =\ \left[ 1\ -\ {\sigma_x^2 \over 2 N^2} t^2\
 +\ {\cal O} \left( {t^3 \over N^3} \right) \right]^N\
 =\ \exp \left[ - {1\over 2}\,{\sigma_x^2 \over N}\, t^2 \right]
 +\ {\cal O} \left( {t^3 \over N^2} \right) \ .$$
This is the {\bf central limit theorem:} The probability density of 
the arithmetic mean $\overline{x}^{\,r}$ converges towards the Gaussian 
probability density with variance (compare Eq.~(\ref{chfct_gau}))
\begin{equation} \label{mean_var}
 \sigma^2 (\overline{x}^{\,r})\ =\ {\sigma^2 (x^r) \over N}\ . 
\end{equation}

{\bf Binning:}
The notion of binning introduced here should not be confused with 
histogramming. Binning means here that we group {\tt NDAT} data 
into {\tt NBINS} bins, where each binned data point is the arithmetic 
average of
$$ {\tt NBIN = [NDAT/NBINS]} ~~({\rm Fortran\ integer\ division})$$
data points in their original order. Preferably {\tt NDAT} is a multiple 
of {\tt NBINS}.  The purpose of the binning procedure is twofold:

\begin{enumerate}
\item When the the central limit theorem applies, the binned data
will become practically Gaussian, as soon as {\tt NBIN} becomes large
enough. This allows to apply Gaussian error analysis methods even
when the original data are not Gaussian.

\item When data are generated by a Markov process subsequent events 
are correlated. For binned data these correlations are reduced
and can in practical applications be neglected, once {\tt NBIN} 
is sufficiently large compared to the autocorrelation time (see
section~\ref{AutoStat}).
\end{enumerate}

\section{Gaussian Error Analysis for Large and Small Samples}

The central limit theorem underlines the importance of the normal 
distribution. Assuming we have a large enough sample, the 
arithmetic mean of a suitable expectation value becomes normally 
distributed and the calculation of the confidence intervals is reduced 
to studying the normal distribution. It has become the convention to 
use the {\bf standard deviation} of the sample mean
\begin{equation} \label{mean_sdev}
 \sigma = \sigma (\overline{x}^{\,r}) ~~{\rm with}~~~
 \overline{x}^{\,r} = \frac{1}{N} \sum_{i=1}^N x_i^r
\end{equation}
to indicate its confidence intervals
$[{\widehat x}-n\sigma ,{\widehat x}+n\sigma ]$ (the dependence of 
$\sigma$ on $N$ is suppressed). For a Gaussian distribution 
equation Eq.~(\ref{gdf}) yields the probability content $p$ of 
the confidence intervals~(\ref{mean_sdev}) to be
\begin{equation} \label{p_gauss}
 p = p(n) = G(n\sigma) - G(-n\sigma) = 
  {1\over \sqrt{2\pi}} \int_{-n}^{+n} dx\, e^{-{1\over 2} x^2}
  = {\rm erf} \left( {n\over \sqrt{2}} \right)\ .
\end{equation}
In practice the roles of $\overline{x}$ and $\widehat{x}$ are 
interchanged: One would like to know the likelihood that the 
{\bf unknown} exact expectation value ${\widehat x}$ will be 
in a certain confidence interval around the measured sample mean. 
The relationship
\begin{equation} \label{interchange}
 {\overline x}\in [{\widehat x}-n\sigma ,{\widehat x}+n\sigma]\
 \Longleftrightarrow\ {\widehat x} \in
 [ {\overline x}-n\sigma ,{\overline x}+n\sigma ]
\end{equation}
solves the problem. Conventionally, these estimates are quoted as
\begin{equation} \label{error_bar}
 {\widehat x}\ =\ {\overline x} \pm \triangle {\overline x}
\end{equation}
where the \index{error bar}{\bf error bar} $\triangle \overline{x}$ 
is often an {\bf estimator} of the exact standard deviation.

An obvious estimator for the variance 
$\sigma^2_x$ is
\begin{equation} \label{sample_variance1}
  (s'^{\,r}_x)^2\ =\ {1\over N} \sum_{i=1}^N (x_i^r
  - \overline{x}^{\,r})^2
\end{equation}
where the prime indicates that we shall not be happy with it, because
we encounter a {\bf bias}. An estimator is said to be biased when its 
expectation value does not agree with the exact result. In our case
\begin{equation} \label{variance_bias}
 \langle (s'^{\,r}_x)^2\rangle\ \ne\ \sigma^2_x\ . 
\end{equation}
An estimator whose expectation value agrees with the true expectation 
value is called {\bf unbiased}. The bias of the definition 
(\ref{sample_variance1}) comes from replacing the exact mean 
${\widehat x}$ by its estimator $\overline{x}^{\,r}$. The latter 
is a random variable, whereas the former is just a number. Some algebra 
\cite{Berg} shows that the desired {\bf unbiased estimator of the 
variance} is given by
\begin{equation} \label{sample_variance}
 (s^r_x)^2\ =\ {N\over N-1}\, (s'^{\,r}_x)^2\ =\
{1\over N-1} \sum_{i=1}^N ( x_i^r - \overline{x}^{\,r} )^2\ .
\end{equation}
Correspondingly, the unbiased estimator of the variance of the 
sample mean is 
\begin{equation} \label{sample_mean_variance}
 (s^r_{\overline x})^2\ =\ {1\over N (N-1)} 
 \sum_{i=1}^N ( x_i^r - \overline{x}^{\,r} )^2\ .
\end{equation}

{\bf Gaussian difference test:}
In practice one is often faced with the problem to compare two 
different empirical estimates of some mean. How large must 
$D = {\overline x} - {\overline y}$ be in order to indicate a 
real difference? The quotient 
\begin{equation} \label{gdr}
  d^r\ =\ {D^r\over \sigma_D}\,,\qquad \sigma_D =
  \sqrt{\sigma^2_{\overline{x}}+\sigma^2_{\overline{y}}}
\end{equation}
is normally distributed with expectation zero and variance one,
so that
\begin{equation} \label{Pgd}
 P\ =\ P(|d^r| \le d)\ =\ G_0(d) - G_0(-d)\ 
    =\ {\rm erf} \left( {d\over\sqrt{2}} \right)\ .
\end{equation}
The {\bf likelihood that the observed difference 
$|{\overline x}-{\overline y}|$ is due to chance}
is defined to be
\begin{equation} \label{Qgood}
Q\ =\ 1-P\ =\ 2\,G_0(-d)\ =\ 
1-{\rm erf} \left( {d\over\sqrt{2}}\right)\ .
\end{equation}
If the assumption is correct, then $Q$ is a 
uniformly distributed random variable in the range $[0,1)$.
Examples are collected in table~\ref{tab_gdt}. Often a 5\% cut-off
is used to indicate a real discrepancy.

\begin{table}[ht]
\tbl{Gaussian difference tests (compile and run the program 
provided in {\tt ForProc/Gau\_dif}, which results in an interactive
dialogue).} 
{ \begin{tabular}{||c|c|c|c|c|c||}                        \hline
 ${\overline x}_1 \pm \sigma_{{\overline x}_1}$ & $1.0\pm 0.1$
 & $1.0\pm 0.1$ &$1.0\pm 0.1$&$1.0\pm 0.05$ & $1.000\pm 0.025$
\\ \hline
 ${\overline x_2} \pm \sigma_{{\overline x}_2}$ & $1.2\pm 0.2$
 &$1.2\pm 0.1$ &$1.2\pm 0.0$&$1.2\pm 0.00$ & $1.200\pm 0.025$
\\ \hline
 $Q$ & 0.37 & 0.16 &~~~0.046 & 0.000063 & $0.15\times 10^{-7}$
\\ \hline
\end{tabular} \label{tab_gdt} }
\end{table}

{\bf Gosset's Student Distribution:}\index{Student distribution} 
We ask the question: What happens with the Gaussian confidence 
limits when we replace the variance $\sigma^2_{\overline x}$ by 
its estimator $s^2_{\overline x}$ in statements like
$$ { |\overline{x} - \widehat{x} | \over \sigma_{\overline x} }
 < 1.96 ~~{\rm with}~~ 95\% ~~{\rm probability.} $$
For sampling from a Gaussian distribution the answer was given by
Gosset, who published his article 1908 under the pseudonym 
{\it Student} in Biometrika~\cite{Go08}. He showed that the 
distribution of the random variable
\begin{equation} \label{student_rv}
 t^r = { \overline{x}^{\,r} - \widehat{x} \over s^r_{\overline x} }
\end{equation}
is given by the probability density 
\begin{equation} \label{stud_pd}
   f(t) = {1\over (N-1)\, B(1/2,(N-1)/2)}\,
  \left( {1+{t^2\over N-1}} \right)^{-{N\over 2}}\ .
\end{equation} 
Here $B(x,y)$ is the beta function. The fall-off is a power law 
$|t|^{-N}$ for $|t|\to \infty$, instead of the exponential fall-off 
of the normal distribution. Some confidence probabilities of the 
Student distribution are (assignment {\tt a0203\_01}):
\begin{verbatim}  
     N \ S  1.0000     2.0000     3.0000     4.0000     5.0000    
     2      .50000     .70483     .79517     .84404     .87433    
     3      .57735     .81650     .90453     .94281     .96225    
     4      .60900     .86067     .94233     .97199     .98461    
     8      .64938     .91438     .98006     .99481     .99843    
    16      .66683     .93605     .99103     .99884     .99984    
    32      .67495     .94567     .99471     .99963     .99998    
    64      .67886     .95018     .99614     .99983     1.0000    
 INFINITY:  .68269     .95450     .99730     .99994     1.0000    
\end{verbatim}
For $N\le 4$ we find substantial deviations from the Gaussian
confidence levels, whereas up to two standard deviations reasonable 
approximations of Gaussian confidence limits are obtained for 
$N\ge 16$ data. If desired, the Student distribution function
can always be used to calculate the exact confidence limits. 
When the central limit theorem applies, we can bin a large 
set of non-Gaussian data into 16 almost Gaussian data to reduce 
the error analysis to Gaussian methods.

{\bf Student difference test:}
This test is a generalization of the Gaussian difference test. It
takes into account that only a finite number of events are sampled. 
As before it is assumed that the events are drawn from a normal 
distribution.  Let the following data be given
\begin{eqnarray}
\overline{x} ~~{\rm calculated\ from}~~ M ~~{\rm events},\
 {\it i.e.},~~ \sigma^2_{\overline x} &=& \sigma^2_x/M \\
\overline{y} ~~{\rm calculated\ from}~~ N ~~{\rm events},\
 {\it i.e.},~~ \sigma^2_{\overline y} &=& \sigma^2_y/N
\end{eqnarray}
and unbiased estimators of the variances are
\begin{equation} \label{partial_data}
s^2_{\overline x} = s^2_x/M = {\sum_{i=1}^M (x_i-\overline{x})^2
\over M\,(M-1)} ~~{\rm and}~~
s^2_{\overline y} = s^2_y/N = {\sum_{j=1}^N (y_j-\overline{y})^2
\over N\,(N-1)}\ .
\end{equation}
Under the {\bf additional assumption} $\sigma^2_x=\sigma^2_y$ the 
probability 
\begin{equation}
 P(|\overline{x}-\overline{y}|>d)
\end{equation}
is determined by the Student distribution function in 
the same way as the probability of the Gaussian difference test
is determined by the normal distribution. 

Examples for the Student difference test for  
$\overline{x}_1 = 1.00 \pm 0.05$ from $M$ data and 
$\overline{x}_2 = 1.20 \pm 0.05$ from $N$ data
are given in table~\ref{tab_sdt}.
The Gaussian difference test gives $Q=0.0047$. For $M=N=512$ the 
Student $Q$ value is practically identical with the Gaussian result, 
for $M=N=16$ it has almost doubled. Likelihoods above a 5\% cut-off, 
are only obtained for $M=N=2$ (11\%) and $M=16$, $N=4$ (7\%). The 
latter result looks a bit surprising, because its $Q$ value is 
smaller than for $M=N=4$.  The explanation is that for $M=16$, 
$N=4$ data one would expect the $N=4$ error bar to be two times 
larger than the $M=16$ error bar, whereas the estimated error bars 
are identical.
This leads to the problem: Assume data are sampled from the same 
normal distribution, when are two measured error bars consistent 
and when not?

\begin{table}[ht]
\tbl{Student difference test for the data 
$\overline{x}_1 = 1.00 \pm 0.05$ and
$\overline{x}_2 = 1.20 \pm 0.05$
(compile and run the program provided 
in {\tt ForProc/Stud\_dif}, which results in an interactive 
dialogue).}
{ \begin{tabular}{||c|c|c|c|c|c|c|c||}    \hline
$M$& 512  & 32   & 16   & 16  & 4   & 3   & 2  \\
\hline
$N$& 512  & 32   & 16   &  4  & 4   & 3   & 2  \\
\hline
$Q$&0.0048&0.0063&0.0083&0.072&0.030&0.047&0.11\\
\hline
\end{tabular} } \label{tab_sdt}
\end{table}

\subsection{$\chi^2$ Distribution, Error of the Error Bar, F-Test} 

\index{chi squared distribution}The distribution of the random variable
\begin{equation} \label{chi2_r}
 (\chi^r)^2 = \sum_{i=1}^N (y_i^r)^2\ ,
\end{equation}
where each $y_i^r$ is normally distributed, defines the
{\bf ${\bf\chi^2}$ distribution} with $N$ degrees of freedom. 
The study of the variance $(s^r_x)^2$ of a Gaussian sample can be 
reduced to the $\chi^2$-distribution with $f=N-1$ degrees of freedom
\begin{equation} \label{chi2_s2}
 (\chi_f^r)^2 = {(N-1)\, (s^r_x)^2 \over \sigma^2_x } = 
 \sum_{i=1}^N { (x^r_i - \overline{x}^{\,r})^2 \over \sigma^2_x }\ .
\end{equation}
The probability density of {\bf $\chi^2$ per degree of freedom (pdf)} 
is
\begin{equation} \label{chi2_pdf_pd}
 f_N(\chi^2)\ =\ N f(N\chi^2)\ = 
{a\ e^{- a \chi^2}\ \left( a \chi^2 \right)^{a-1} \over
\Gamma (a) } ~~{\rm where}~~ a={N\over 2}\ . 
\end{equation}

{\bf The Error of the Error Bar:}\index{error bar}
For normally distributed data the number of data alone determines
the errors of error bars, because the $\chi^2$ distribution is
exactly known. 
Confidence intervals for variance estimates $s^2_x=1$ 
from {\tt NDAT} data (assignment {\tt a0204\_01}) are:
\begin{verbatim}
                      q         q         q       1-q       1-q

    NDAT=2**K      .025      .150      .500      .850      .975
 
       2    1      .199      .483     2.198    27.960  1018.255
       4    2      .321      .564     1.268     3.760    13.902
       8    3      .437      .651     1.103     2.084     4.142
      16    4      .546      .728     1.046     1.579     2.395
      32    5      .643      .792     1.022     1.349     1.768
    1024   10      .919      .956     1.001     1.048     1.093
   16384   14      .979      .989     1.000     1.012     1.022
\end{verbatim}

{\bf The variance ratio test or F-test:}\index{F-Test}
\index{variance ratio}We assume that two sets of normal data are 
given together with estimates of their variances: 
$ \left( s^2_{x_1},N_1\right)$ and $\left( s^2_{x_2},N_2\right)$. 
We would like to test whether the 
ratio $F={s^2_{x_1}/s^2_{x_2}}$ differs from $F=1$ in a statistically 
significant way. The probability $(f_1/f_2)\,F < w$, where 
$f_i=N_i-1,\ i=1,2$, is known to be
\begin{equation} \label{Hw_beta}
 H(w) = 1 - B_I \left( {1\over w + 1}, {1\over 2}\,f_2,
 {1\over 2}\,f_1 \right)\ .
\end{equation}
Examples are given in table~\ref{tab_Ftest}. This allows us later to 
compare the efficiency of MC algorithms.
 
\begin{table}[ht]
\tbl{Examples for the F-test (use the program in 
{\tt ForProc/F\_test} or the one in {\tt ForProc/F\_stud}).}
{ \begin{tabular}{||c|c|c|c|c|c|c|c|c||}    \hline
$\triangle \overline{x}_1$
     & 1.0& 1.0&  1.0 &1.0 & 1.0 &1.0&  1.0& 1.0\\
\hline
$N_1$&  16&  16&   64 &1024& 2048& 32& 1024& 16 \\
\hline
$\triangle \overline{x}_2$
     & 1.0& 1.0&  1.0 &1.05& 1.05&2.0&  2.0& 2.0\\
\hline
$N_2$& 16 &   8&   16 &1024& 2048&  8&  256& 16 \\
\hline
$Q$&  1.0& 0.36& 0.005&0.12&0.027&0.90&0.98&0.01\\
\hline
\end{tabular} } \label{tab_Ftest}
\end{table}

\subsection{The Jackknife Approach} 

Jackknife estimators allow to correct for the bias and the error of 
the bias. The method was introduced in the 1950s (for a review 
see~\cite{Berg}).  It is {\bf recommended as the standard} for error 
bar calculations. In unbiased situations the jackknife and the usual 
error bars agree. Otherwise the jackknife estimates are improvements. 

The unbiased estimator of the expectation value $\widehat{x}$ is
$$ {\overline x} = {1\over N} \sum_{i=1}^N x_i $$
Bias problems may occur when one estimates a non-linear function 
of ${\widehat x}$: 
\begin{equation} 
 {\widehat f} = f({\widehat x})\ . 
\end{equation}
Typically, the bias is of order $1/N$:
\begin{equation} \label{f_bias}
 {\rm bias}\ ({\overline f})\ =\ {\widehat f} -  
 \langle {\overline f}\rangle\ =\ {a_1 \over N} + {a_2 \over N^2} 
 + O({1\over N^3})
\end{equation}
where $a_1$ and $a_2$ are constants. But for the biase estimator we 
lost the ability to estimate the variance $\sigma^2 ({\overline f}) 
= \sigma^2 (f) / N$ via the standard equation
\begin{equation} \label{f_bad_variance}
 s^2 ({\overline f})\ =\ {1\over N} s^2 (f)\ =\
 {1\over N\, (N-1)} \sum_{i=1}^N (f_i - {\overline f})^2\ ,  
\end{equation}
because $f_i=f(x_i)$ is not a valid estimator of ${\widehat f}$. 
Further, it is in non-trivial applications almost always a bad idea 
to use linear error propagation formulas. Jackknife methods are not 
only easier to implement, but also more precise and far more 
{\bf robust}.

The error bar problem for the estimator $\overline{f}$ is conveniently 
overcome by using {\bf jackknife estimators} ${\overline f^J}$, 
$f_i^J$, defined by
\begin{equation} \label{jackknife_estimators}
 {\overline f^J}\ =\ {1\over N} \sum_{i=1}^N f^J_i
 ~~{\rm with}~~ f_i^J\ =\ f(x_i^J) ~~{\rm and}~~ 
   x_i^J\ =\ {1\over N-1} \sum_{k\ne i} x_k\ . 
\end{equation}
The estimator for the variance $\sigma^2 ({\overline f^J})$ is
\begin{equation} \label{jackknife_variance}
 s^2_J ({\overline f^J})\ =\ {N-1 \over N} 
 \sum_{i=1}^N (f^J_i - {\overline f^J})^2\ . 
\end{equation}
Straightforward algebra shows that in the unbiased case the estimator 
of the jackknife variance~(\ref{jackknife_variance}) reduces to the 
normal variance~(\ref{f_bad_variance}). 
Notably only of order $N$ (not $N^2$) operations are needed to
construct the jackknife averages $x_i^J,\ i=1,\dots ,N$ from the
orginal data.

 
\section{Statistial Physics and Potts Models} 

MC simulations of systems described by the \index{Gibbs}Gibbs 
\index{canonical ensemble}{canonical ensemble} 
aim at calculating estimators of physical observables at a 
temperature $T$. In the following we choose units so that the 
\index{Boltzmann}Boltzmann 
constant becomes one, i.e. $\beta = 1/T$. Let us consider 
the calculation of the {\bf expectation value} of an {\bf observable} 
$\mathcal{O}$. Mathematically all systems on a computer are discrete, 
because a finite word length has to be used. Hence, the expectation 
value is given by the sum
\begin{eqnarray} \label{O}
 \widehat{\mathcal{O}} = \widehat{\mathcal{O}} (\beta) = 
 \langle \mathcal{O} \rangle &=& Z^{-1} \sum_{k=1}^K
 \mathcal{O}^{(k)}\,e^{-\beta\,E^{(k)} }\\
{\rm where}~~~
 Z\ =\ Z(\beta) &=& \sum_{k=1}^K e^{-\beta\,E^{(k)} }
\end{eqnarray}
is the {\bf partition function}. The index $k=1,\dots , K$ labels the 
{\bf configurations} of the system, and $E^{(k)}$ is the (internal) 
{energy } of configuration $k$. The configurations are also called 
{\bf microstates}.  To distinguish the configuration index 
from other indices, it is put in parenthesis.

We introduce generalized \index{Potts models}Potts models in an  
external magnetic field on $d$-dimensional hypercubic lattices with 
periodic boundary conditions (i.e., the models are defined on a
torus in $d$ dimensions). 
Without being overly complicated, these models are general
enough to illustrate the essential features we are interested in.
In addition, various subcases of these models are by themselves of 
physical interest. 

We define the energy function of the system by
\begin{equation} \label{E}
 -\beta\,E^{(k)} = -\beta\,E_0^{(k)} + H\,M^{(k)} 
\end{equation}
where
\begin{equation} \label{E0}
  E_0^{(k)} = - 2\,\sum_{\langle ij\rangle} \delta(q_i^{(k)},q_j^{(k)}) 
           + {2\,d\,N\over q}
\end{equation}
$$ {\rm with}~~\ \delta (q_i,q_j)
 = \left\{ \begin{array}{c} 1\ {\rm for}\ q_i=q_j \\
         0\ {\rm for}\ q_i\ne q_j \end{array}  \right. 
~~~{\rm and}~~~
 M^{(k)} = 2 \sum_{i=1}^N \delta (1,q_i^{(k)})\, .  $$

The sum $\langle ij\rangle$ is over the nearest neighbor lattice sites 
and $q_i^{(k)}$ is called the {\bf Potts spin} or {\bf Potts state} of 
configuration $k$ at site $i$. For the $q$-state Potts model $q^{(k)}_i$ 
takes on the values $1,\dots ,q$. The external 
\index{magnetization}magnetic field is chosen 
to interact with the state $q_i=1$ at each site $i$, but not with the 
other states $q_i\ne 1$. The case $q=2$ becomes equivalent to the 
Ising ferromagnet. See F.Y. Wu \cite{Wu82} for a detailed review of
Potts models. 

For the \index{energy}{\bf energy per spin} our notation is
\begin{equation} \label{Potts_es}
 e_s = E/N\ .
\end{equation}
A factor of two and an additive constant are introduced in 
Eq.~(\ref{E0}), so that $e_s$ agrees 
for $q=2$ with the conventional Ising model definition, and
\begin{equation} \label{beta_Potts}
\beta\ =\ \beta^{\rm Ising}\ =\ {1\over 2}\, \beta^{\rm Potts}\, .
\end{equation}
For the $2d$ Potts models a number of exact results are known in the
infinite volume limit, mainly due to work by Baxter~\cite{Ba73}. The 
phase transions temperatures are
\begin{equation} \label{2d_Potts_bc}
{1\over 2}\, \beta_c^{\rm Potts}\ =\ \beta_c\ =\ {1\over T_c}\ 
=\ {1\over 2}\, \ln (1 + \sqrt{q} ), ~~q=2,3,\dots\ .
\end{equation}
At $\beta_c$ the average energy per state is 
\begin{equation} \label{es_bc}
 e_s^c = E_0^c/N = {4\over q} - 2 - 2/\sqrt{q}\ .
\end{equation}
The phase transition is second order for $q\le 4$ and first order 
for $q\ge 5$. The exact infinite volume {\bf latent heats} 
$\triangle e_{s}$ and {\bf entropy jumps} $\triangle s$ were also 
found by Baxter \cite{Ba73}, while the interface tensions $f_s$ were 
derived later (see \cite{BoJa92} and references therein).

\section{Sampling and Re-weighting}

\begin{figure}[tb]
\vspace{5pc}
 \centerline{\hbox{ \psfig{figure=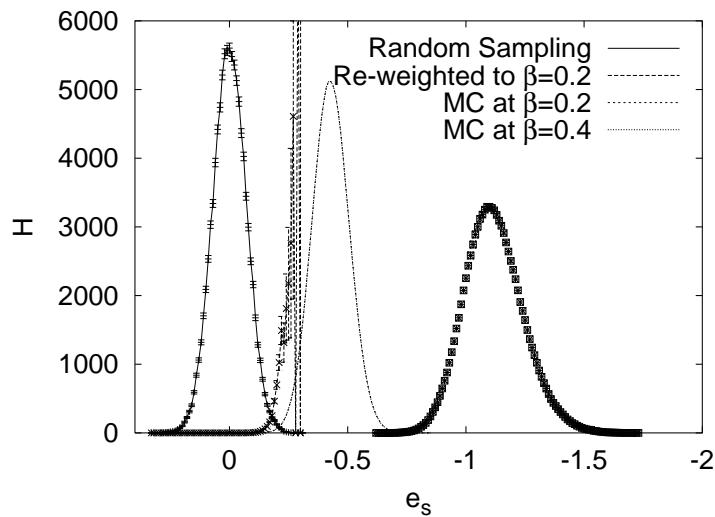,width=10cm} }}
\caption{Energy histograms of 100$\,$000 entries each for the Ising 
model on a $20\times 20$ lattice: Random Sampling gives statistically 
independent configurations at $\beta=0$. Histograms at $\beta=0.2$ 
and $\beta=0.4$ are generated with Markov chain MC. Re-weighting of 
the $\beta=0$ random configurations to $\beta=0.2$ is shown to fail 
(assignments {\tt a0301\_02} and {\tt a0303\_02}).}
 \label{fig_2dI_h_es}
\end{figure}

For the Ising model it is straightforward to {\bf sample statistically 
independent configurations}. We simply have to generate $N$ spins, each 
either up or down with 50\% likelihood. This is called {\bf random 
sampling}. In Fig.~\ref{fig_2dI_h_es} a thus obtained histogram for 
the $2d$ Ising model {\bf energy per spin} is depicted. 

Note that is is very important to distinguish the energy measurements 
on single configurations from the expectation value. The expectation 
value $\widehat{e}_{s}$ is a single number, while $e_{s}$ fluctuates.
From the measurement of many $e_{s}$ values one finds an estimator
of the mean, $\overline{e}_{s}$, which fluctuates too.

The histogram entries at $\beta=0$ can be re-weighted so that they 
correspond to other $\beta$ values. We simply have to multiply the 
entry corresponding to energy $E$ by $\exp (-\beta E)$.
Similarly histograms corresponding to the Gibbs ensemble at some 
value $\beta_0$ can be re-weighted to other $\beta$ values. 
Care has to be taken to ensure that the involved arguments of the 
exponential function do not become too large. This can be done by 
first calculating the mean energy and then implementing 
re-weighting with respect to the difference from the mean. 

Re-weighting has a long history. For finite size scaling (FSS) 
investigations of second order phase transitions its usefulness has 
been stressed by Ferrenberg and Swendsen \cite{FeSw88} (accurate 
determinations of peaks of the specific heat or of susceptibilities). 
 
In Fig.~\ref{fig_2dI_h_es} re-weighting is done from $\beta_0=0$ to
$\beta=0.2$. But, by comparison to the histogram from a Metropolis 
MC calculation at $\beta=0.2$, the result is seen to be disastrous. 
The reason is easily identified: In the range where the $\beta=0.2$ 
histogram takes on its maximum, the $\beta=0$ histogram  has not a 
single entry. Our random sampling procedure misses the important 
configurations at $\beta =0.2$. Re-weighting to new $\beta$ values 
works only in a range $\beta_0\pm \triangle \beta$, where 
$\triangle\beta\to 0$ in the infinite volume limit. 

{\bf Important Configurations:}
Let us determine the important contributions to the partition function. 
The partition function can be re-written as a sum over energies
\begin{equation} \label{Z_sum_E}
 Z = Z(\beta) = \sum_E n(E)\, e^{-\beta\,E}
\end{equation}
where the unnormalized spectal density $n(E)$ is defined as the number 
of microstates $k$ with energy $E$. For a fixed value of $\beta$ the 
energy probability density
\begin{equation} \label{P_E}
 P_{\beta} (E) = c_{\beta}\, n(E)\, e^{-\beta E} 
\end{equation}
is peaked around the average value $\widehat{E}(\beta)$, where 
$c_{\beta}$ is a normalization constant determined by
$\sum_E P_{\beta}(E)=1$. 

Away from first and second order phase transitions, the width of the 
energy distribution is $\triangle E\sim\sqrt{V}$. This follows from 
the fact that the fluctuations of the $N\sim V$ lattice spins are 
essentially uncorrelated, so that the magnitude of a typical 
fluctuations is $\sim \sqrt{N}$. As the energy is an extensive 
quantity $\sim V$, we find that the re-weighting range is
$\triangle \beta \sim 1/\sqrt{V}$, so that 
$\triangle \beta E\sim \sqrt{V}$ stays within 
the fluctuation of the system. 

Interestingly, the re-weighting range increases at a second order 
phase transition point, because critical fluctuations are larger than 
non-critical fluctuations. Namely, one has $\triangle E \sim V^x$ 
with $1/2<x<1$ and the requirement $\triangle\beta E\sim V^x$ yields
$\triangle\beta \sim V^{x-1}$. 
 
For first order phase transitions
one has a latent heat $\triangle V\sim V$, but this does not mean
that the re-weighting range becomes of order one. In essence, the 
fluctuations collapse, because the two phases become separated by
an interface. One is back to fluctuations within either of the two 
phases, {\it i.e.} $\triangle\beta\sim 1/\sqrt{V}$.

The important 
configurations at temperature $T=1/\beta$ are at the energy values for
which the probability density $P_{\beta}(E)$ is large. To sample them 
efficiently, one needs a procedure which generates the configurations 
with their \index{Boltzmann}Boltzmann \index{weights}weights
\begin{equation} \label{w_B}
 w_B^{(k)} = e^{-\beta E^{(k)}}\ . 
\end{equation}
The number of configurations $n(E)$ and the weights combine then 
so that the probability to generate a configuration at energy $E$
becomes precisely $P_{\beta}(E)$ as given by equation~(\ref{P_E}).

\section{Importance Sampling and Markov Chain Monte Carlo}

For the canonical ensemble {\bf importance sampling} generates 
configurations $k$
with probability
\begin{equation} \label{P_B}
 P_B^{(k)} = c_B\, w^{(k)}_B = c_B\, e^{-\beta E^{(k)}}
\end{equation}
where the constant $c_B$ is determined by the normalization condition
$\sum_k P_B^{(k)} = 1$. The vector $(P_B^{(k)})$ is called 
{\bf Boltzmann state}. When configurations are stochastically 
generated with probability $P_B^{(k)}$, the {\bf expectation value} 
becomes the {\bf arithmetic average:}
\begin{equation} \label{O_B}
  {\widehat {\cal O}} = {\widehat {\cal O}} (\beta) =
  \langle {\cal O} \rangle  = \lim_{N_K\to \infty} 
  {1\over N_K} \sum_{n=1}^{N_K} {\cal O}^{(k_n)}\ .
\end{equation}
Truncating the sum at some finite value of $N_K$, we obtain an 
\index{expectation value}{\bf estimator of the expectation 
value}
\begin{equation} \label{estimate_O_B}
 {\overline {\cal O}} =
 {1\over N_K} \sum_{n=1}^{N_K} {\cal O}^{(k_n)}\ .
\end{equation}
Normally, we cannot generate configurations $k$ directly with the
probability~(\ref{P_B}), but they may be found as members
of the equilibrium distribution of a dynamic process. A 
\index{Markov process}{\bf Markov process} is a particularly
simple dynamic process, which generates configuration $k_{n+1}$
stochastically from configuration $k_n$, so that no information
about previous configurations $k_{n-1}, k_{n-2}, \dots$ is needed.
The elements of the Markov process
\index{time series}{\bf time series} are the configurations.
Assume that the configuration $k$ is given. Let the 
\index{transition probability}{transition probability}
to create the configuration $l$ in one step from $k$ be 
given by $W^{(l)(k)} = W[k\to l]$. The 
\index{transition matrix}{\bf transition matrix}
\begin{equation} \label{W}
 W = \left( W^{(l)(k)}\right)
\end{equation}
defines the Markov process. Note, that this matrix is a very big (never 
stored in the computer), because its labels are the configurations. To 
generate configurations with the desired probabilities, the matrix $W$ 
needs to satisfy the following properties:

\begin{description}

\item{(i)} \index{ergodicity}{\bf Ergodicity}:
\begin{equation} \label{ergodicity}
  e^{-\beta E^{(k)}} > 0 ~~{\rm and}~~ e^{-\beta E^{(l)}} > 0
  ~~~{\rm imply:}
\end{equation}
an integer number $n>0$ exists so that $(W^n)^{(l)(k)}>0$ holds.

\item{(ii)} {\bf Normalization}:
\begin{equation} \label{normalization}
 \sum_l W^{(l)(k)} = 1\ .
\end{equation}

\item{(iii)} \index{balance}{\bf Balance}: 
\begin{equation} \label{balance}
 \sum_k W^{(l)(k)}\, e^{-\beta E^{(k)}}\ =\ e^{-\beta E^{(l)}}\ .
\end{equation}
Balance means: The Boltzmann state~(\ref{P_B}) is an
eigenvector with eigenvalue $1$ of the matrix $W = (W^{(l)(k)})$.

\end{description}

An \index{ensemble}{\bf ensemble} is a collection of configurations for 
which to each configuration $k$ a probability $P^{(k)}$ is assigned, 
$\sum_k P^{(k)} = 1$. The \index{Gibbs}{\bf Gibbs or Boltzmann ensemble} 
\index{Boltzmann}$E_B$ is defined to be the ensemble with the probability 
distribution~(\ref{P_B}). 

An {\bf equilibrium ensemble} $E_{eq}$ of the Markov process is defined 
by its probability distribution $P_{eq}$ satisfying
\begin{equation} \label{E_eq}
 W\, P_{eq} = P_{eq}\, , ~~{\rm in\ components}~~
 P_{eq}^{(l)}=\sum_k W^{(l)(k)} P_{eq}^{(k)}\ .
\end{equation}

{\bf Statement:} Under the conditions (i), (ii) and (iii)
the Boltzmann ensemble is the {\bf only} equilibrium ensemble of the
Markov process.

For a proof the readers is referred to~\cite{Berg}. There are 
many ways to construct a Markov process satisfying (i), (ii)
and (iii).  A stronger condition than balance~(\ref{balance}) is

\begin{description}

\item{(iii')} \index{balance}{\bf Detailed balance}:
\begin{equation} \label{detailed_balance}
 W^{(l)(k)}\, e^{-\beta E^{(k)}}\ =\ W^{(k)(l)} e^{-\beta E^{(l)}}\ .
\end{equation}
Using the normalization $\sum_k W^{(k)(l)} = 1$ detailed
balance implies balance (iii).

\end{description}

At this point we have succeeded to replace the canonical ensemble
average by a time average over an artificial dynamics. Calculating
averages over large times, like one does in real experiments,
is equivalent to calculating averages of the ensemble. One 
distinguishes \index{dynamics}{\it dynamical universality classes}.  
The Metropolis and heat bath algorithms discussed in the following
fall into the class of so called {\it Glauber dynamics}, model~A
in a frequently used classification~\cite{ChLu97}. Cluster 
algorithms \cite{SwWa87} constitute another universality class. 

\subsection{Metropolis and Heat Bath Algorithm for Potts Models}

The \index{Metropolis algorithm|textbf}{\bf Metropolis algorithm} 
can be used whenever one knows how to calculate the energy of a 
configuration. Given a configuration $k$, the Metropolis algorithm 
proposes a configuration $l$ with probability
\begin{equation} \label{Metropolis_f}
 f(l,k) ~~{\rm normalized\ to}~~~ \sum_l f(l,k)=1\ .
\end{equation}
The new configuration $l$ is accepted with probability
\begin{equation} \label{Metropolis}
 w^{(l)(k)} = \min \left[ 1,\, {P_B^{(l)}\over P_B^{(k)}} \right] 
 = \left\{ 
   \begin{matrix} 1& {\rm for}& E^{(l)}<E^{(k)}\\
   e^{-\beta (E^{(l)}-E^{(k)})} &{\rm for}& E^{(l)}>E^{(k)}.
   \end{matrix} \right.
\end{equation}
If the new configuration is rejected, the old configuration has to 
be counted again. 
The {\bf acceptance rate} is defined as the ratio of accepted changes 
over proposed moves. With this convention we do not count a move as 
accepted when it proposes the at hand configuration.

The Metropolis procedure gives rise to the transition probabilities 
\begin{eqnarray} \label{Metropolis_W_ne}
 W^{(l)(k)} &=& f(l,k)\, w^{(l)(k)}\ \ {\rm for}\ \ l \ne k\\
{\rm and}~~
 W^{(k)(k)}& =& f(k,k) + \sum_{l\ne k} f(l,k)\, (1-w^{(l)(k)})\ .
\end{eqnarray}
Therefore, the ratio $\left( W^{(l)(k)}/W^{(k)(l)} \right)$ satisfies 
detailed balance~(\ref{detailed_balance}) if
\begin{equation} \label{symmetry}
 f(l,k)\ =\ f(k,l) ~~{\rm holds}\,.
\end{equation}
Otherwise the probability density $f(l,k)$ is unconstrained. So there 
is an amazing flexibility in the choice of the transition probabilities 
$W^{(l)(k)}$.  Also, the algorithm generalizes 
immediately to arbitrary weights.

The \index{heat bath algorithm}{\bf heat bath algorithm} 
chooses a state $q_i$ directly with the local Boltzmann distribution 
defined by its nearest neighbors. The state $q_i$ can take on one of 
the values $1,\dots,q$ and, with all 
other states set, determines a value of the energy function~(\ref{E}). 
We denote this energy by $E(q_i)$ and the Boltzmann probabilities are
\begin{equation} \label{Potts_B_weights}
 P_B(q_i)\ =\ {\rm const}\ e^{-\beta\, E(q_i)}
\end{equation}
where the constant is determined by the normalization condition
\begin{equation} \label{Potts_B_normalization}
 \sum_{q_i=1}^q P_B(q_i)\ =\ 1\ .
\end{equation}        
In equation~(\ref{Potts_B_weights}) we can define $E(q_i)$ to be just 
the contribution of the interaction of $q_i$ with its nearest neighbors 
to the total energy and absorb the other contributions into the overall 
constant. Here we give a generic code which works for arbitrary values 
of $q$ and $d$ (other implementations may be more efficient).

We calculate the cumulative distribution function of the 
heat bath probabilities
\begin{equation} \label{prob_hb}
P_{HB}(q_i) = \sum_{q'_i=1}^{\tt q_i} P_B(q'_i)\ .
\end{equation}
The normalization condition~(\ref{Potts_B_normalization}) implies
$P_{HB}(q)=1$. Comparison of these cumulative probabilities
with a uniform random number $x^r$ yields the heat bath update
$q_i\to q'_i$. Note that in the heat bath procedure the original value
$q_i^{\rm in}$ does not influence the selection of $q_i^{\rm new}$. 

\subsection{The $O(3)\ \sigma$ Model and the Heat Bath Algorithm}

We give an example of a model with a continuous energy function. 
Expectation values are calculated with respect to the partition function
\begin{equation} \label{O3_Z}
 Z\ =\ \int \prod_i ds_i\ e^{-\beta E(\lbrace s_i \rbrace )}\ .
\end{equation}
\begin{equation} \label{O3_spin}
{\rm The\ spins}~~ \vec{s}_i = 
  \left( \begin{matrix} s_{i,1} \\ s_{i,2} \cr s_{i,3} 
         \end{matrix} \right) 
   ~~{\rm are\ normalized\ to}~~ (\vec{s}_i)^2=1
\end{equation}
\begin{equation} \label{O3_measure}
{\rm and\ the\ measure}\ ds_i\ {\rm is\ defined\ by}\
 \int ds_i\ =\ {1\over 4\pi} \int_{-1}^{+1} d \cos (\theta_i) 
                              \int_0^{2\pi}  d \phi_i\ , 
\end{equation}
where the polar ($\theta_i$) and azimuth ($\phi_i$) angles define the 
spin $s_i$ on the unit sphere. The energy is
\begin{equation} \label{O3_energy}
 E\ =\ - \sum_{\langle ij\rangle} \vec{s}_i \vec{s}_j\ ,  
\end{equation}
where the sum goes over the nearest neighbor sites of the lattice and
$vec{s}_i \vec{s}_j$ is the dot product of the vectors. 
The $2d$ version of the model is of interest to field theorists because 
of its analogies with the four-dimensional Yang-Mills theory. In 
statistical physics the $d$-dimensional model is known as 
the\index{O(3) $\sigma$-model}
\index{Heisenberg ferromagnet}{\bf Heisenberg ferromagnet}
(references can be found in \cite{Berg}). 

We would like to update a single spin $\vec{s}$. The sum of its $2d$ 
neighbors is
$$\vec{S} = \vec{s}_1+\vec{s}_2+\dots +\vec{s}_{2d-1}+\vec{s}_{2d}\ .$$
Hence, the contribution of spin $\vec{s}$ to the energy is
$2d-\vec{s} \vec{S}$. We propose a new spin $\vec{s}^{\,'}$ with the
measure~(\ref{O3_measure}) by drawing two uniformly distributed random 
numbers
\begin{eqnarray} \nonumber
  \phi^r &\in & [0,2\pi) ~~~{\rm for\ the\ azimuth\ angle\ and} \\
  \cos (\theta^r) &=& x^r \in [-1,+1) ~~~{\rm for\ the\ cosine\ of\
   the\ polar\ angle}\,. \nonumber
\end{eqnarray}
This defines the probability function $f(\vec{s}^{\,'},\vec{s})$ of 
the Metropolis process, which accepts the proposed spin
$\vec{s}^{\,'}$ with probability 
$$ w(\vec{s}\to \vec{s}^{\,'})\ =\ \left\{ \begin{matrix} 
 1 & {\rm for} & \vec{S} \vec{s}^{\,'} > \vec{S} \vec{s}\, , \\
 e^{ -\beta (\vec{S} \vec{s} - \vec{S}\vec{s}^{\,'}) } & {\rm for} &
 \vec{S}\vec{s}^{\,'} < \vec{S}\vec{s}\, . 
 \end{matrix} \right.    $$

If sites are chosen with the uniform probability distribution $1/N$ 
per site, where $N$ is the total number of spins, it is obvious that 
the algorithm fulfills detailed balance. It is noteworthy that the 
procedure remains valid when the spins are chosen in the systematic 
order $1,\dots ,N$.  Balance~(\ref{balance}) still holds, whereas 
detailed balance~(\ref{detailed_balance}) is violated (an exercise 
of Ref.~\cite{Berg}). 

One would prefer to choose $\vec{s}^{\,'}$ directly with the
probability
$$ W(\vec{s}\to \vec{s}^{\,'})\ =\ P(\vec{s}^{\,'};\vec{S})\ =\
  {\rm const}\, e^{\beta\, \vec{s}^{\,'} \vec{S}}\ . $$
The \index{heat bath algorithm}{\bf heat bath algorithm} creates this 
distribution.  Implementation of it becomes feasible when the energy 
function allows for an explicit calculation of the probability 
$P(\vec{s}^{\,'};\vec{S})$. This is an easy task for the 
$O(3)$ $\sigma$-model. Let
$$ \alpha = {\rm angle} (\vec{s}^{\,'},\vec{S}),\ \
x=\cos (\alpha )\ \ {\rm and}\ \  S=\beta |\vec{S}|\ . $$
For $S=0$ a new spin $\vec{s}^{\,'}$ is simply obtained by random
sampling. We assume in the following $S>0$. 
The Boltzmann weight becomes $\exp (x S)$ and the normalization
constant follows from
$$ \int_{-1}^{+1} dx\, e^{x S}\ =\ {2\over S}\, \sinh (S)\ .$$
Therefore, the desired probability is
$$ P(\vec{s}^{\,'};\vec{S})\ =\  
  {S \over 2\sinh (S)}\, e^{x S}\ =: f(x) $$
and the method of Eq.~(\ref{gd}) can be used to generate events 
with the probability density $f(x)$. A uniformly distributed random 
number $y^r\in [0,1)$ translates into
\begin{equation} \label{O3_heat_bath-xr}
  x^r = \cos{\alpha^r} = {1\over S} \ln\, [\, \exp (+S) -
  y^r\, \exp (+S) + y^r\, \exp (-S) ]\ . 
\end{equation}
Finally, one has to give $\vec{s}^{\,'}$ a direction in the plane
orthogonal to $S$. This is done by choosing a random angle $\beta^r$
uniformly distributed in the range $0\le\beta^r < 2\pi$. Then,
$x^r=\cos{\alpha^r}$ and $\beta^r$ completely determine
$\vec{s}^{\,'}$ with respect to $\vec{S}$. Before storing
$\vec{s}^{\,'}$ in the computer memory, we have to calculate
coordinates of $\vec{s}^{\,'}$ with respect to a Cartesian
coordinate system, which is globally used for all spins of the
lattice. This amounts to a \index{linear transformation}
\index{transformations, linear}{linear transformation}. 

\subsection{Example Runs}

{\bf Start and equilibration:}\index{equilibration}
Under repeated application of one of our updating procedures the 
probability of states will approach the Boltzmann distribution. 
However, initially we have to start with a microstate which may be 
far off the Boltzmann distribution. Suppression factors like
$10^{-10000}$ are well possible. Although the weight of states 
decreases with $1/n$ where $n$ is the number of steps of the Markov 
process, one should exclude the initial states from the equilibrium 
statistics. In practice this means we should allow for a certain number 
of sweeps {\tt nequi} to equilibrate the system. One {\bf sweep}
updates each spin once or once in the average.

Many ways to generate start configurations exist. Two natural and
easy to implement choices are:

\begin{enumerate}
\item Generate a random configuration corresponding to $\beta =0$.
This defines a {\bf random} or {\bf disordered start} of a MC simulation.

\item Generate a configuration for which all Potts spins take on the 
same $q$-value. This is called an {\bf ordered start} of a MC simulation.
\end{enumerate}

\begin{figure}[tb]
\vspace{5pc}
 \centerline{\hbox{ \psfig{figure=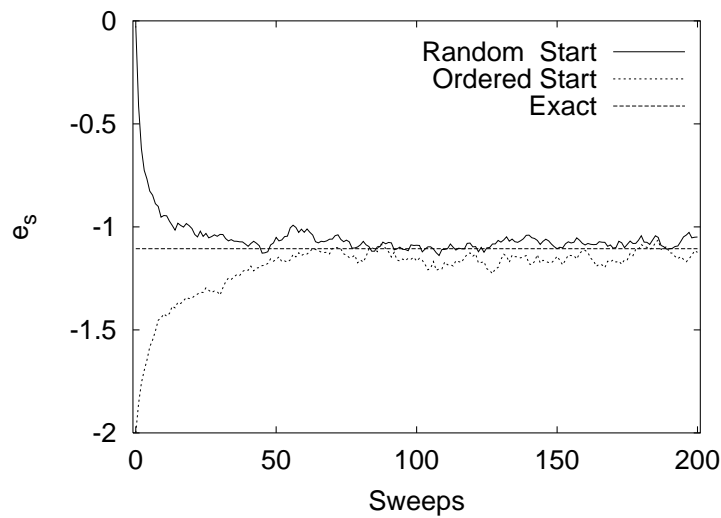,width=10cm} }}
 \caption{Two Metropolis time series of 200 sweeps each for a $2d$ 
  Ising model on a $80\times 80$ lattice at $\beta=0.4$ are shown. 
  Random updating for which the positions of the spins are chose with
  the uniform probability distribution was used. Measurements of the 
  energy per spin after every sweep are plotted for ordered and 
  disordered starts. The exact mean value $\widehat{e}_s=-1.10608$ 
  is also indicated (assignment {\tt a0303\_01}).} \label{fig_2dI_ts}
\end{figure}

\begin{figure}[tb]
\vspace{5pc}
 \centerline{\hbox{ \psfig{figure=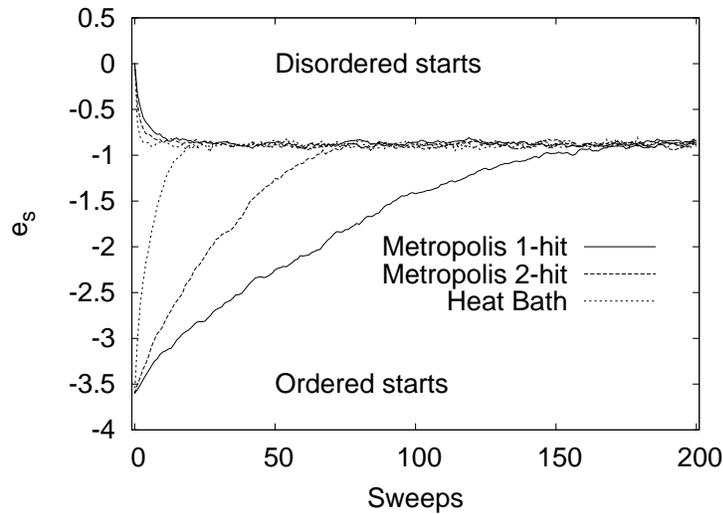,width=10cm} }}
 \caption{$q=10$ Potts model time series of 200 sweeps on a
 $80\times 80$ lattice at $\beta=0.62$. Measurements of the energy
 per spin after every sweep are plotted for ordered and disordered 
 starts (assignment {\tt a0303\_05}). }
 \label{fig_2d10q_ts}
\end{figure}

Examples of initial time series are given in Fig.~\ref{fig_2dI_ts}
and~\ref{fig_2d10q_ts}. Unless explicitly stated otherwise, we use
here and in the following always {\bf sequential updating}, for which 
the spins are touched in a systematic order.
\bigskip

{\bf Consistency Checks:}
For the $2d$ Ising model we can test against the exact finite lattice
results of Ferdinand and Fisher~\cite{FeFi69}. We simulate a $20^2$ 
lattice at $\beta=0.4$, using a statistics of 10$\,$000 sweeps for 
reaching equilibrium. The statistics for measurement is chosen to be 
64 bins of 5$\,$000 sweeps each. The number 64 is taken, because 
according to the student distribution the approximation to the 
Gaussian distribution is then excellent, while the binsize of 5$\,$000 
($\gg 200$) is argued to be large enough to neglect correlations between 
the bins.  A more careful analysis is the subject of our next section.
With our statistics we find (assignment {\tt a0303\_06})
\begin{equation} \label{2dI_es}
\overline{e}_{s} = -1.1172\ (14)~~ {\rm (Metropolis)} 
~~~{\rm versus}~~~ \widehat{e}_s = -1.117834~~ {\rm (exact)}\ .
\end{equation}
The Gaussian difference test gives a perfectly admissible 
value, $Q=0.66$. 

For the $2d$ 10-state Potts model at $\beta =0.62$ we test our 
Metropolis versus our heat bath code on a $20\times 20$ lattice. 
For the heat bath updating we use the same statistics as 
for the $2d$ Ising model. For the Metropolis updating we increase these 
numbers by a factor of four. This increase is done, because we expect 
the performance of Metropolis updating for the 10-state model to be 
worse than for the 2-state model: At low temperature the likelihood 
to propose the most probable (aligned) Potts spin is 1/2 for the 
2-state model, but only 1/10 for the 10-state model, and $\beta=0.62$ 
is sufficiently close to the ordered phase, so that this effect is 
expected to be of relevance. The results of our simulations are
(assignment {\tt a0303\_08}) 
$e_s = -0.88709\ (30)$ (Metropolis) versus
$e_s = -0.88664\ (28)$ (heat\ bath) and $Q=0.27$ for the 
Gaussian difference test. Another perfectly admissable value.

\begin{figure}[tb]
\vspace{5pc}
 \centerline{\hbox{ \psfig{figure=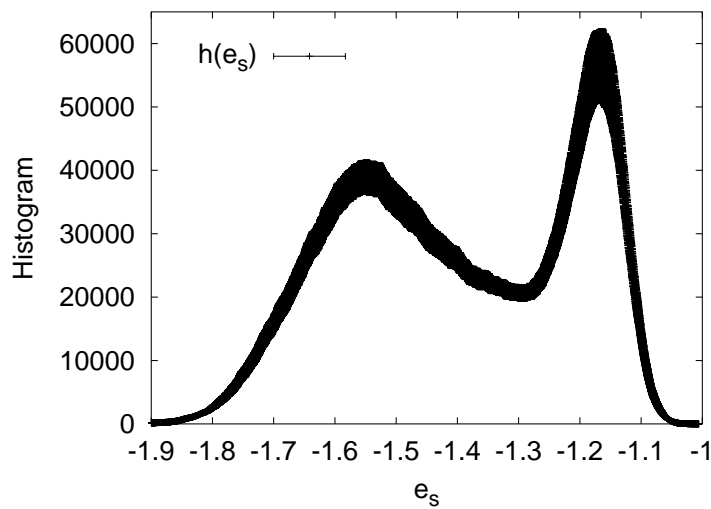,width=10cm} }}
 \caption{Histogram of the energy per spin for the $3d$ 3-state
 Potts model on a $24^3$ latticed at $\beta =0.275229525$
 (assignment {\tt a0303\_10}). } \label{fig_3d3q_h}
\end{figure}

To illustrate features of a first order phase transition for the 
$3d$ 3-state Potts model, we use the 1-hit Metropolis algorithm 
on a $24^3$ lattice and simulate at $\beta = 0.275229525$. We 
perform 20$\,$000 sweeps for reaching equilibrium, then 
$64\times 10$\,$000$ sweeps with measurements. From the latter 
statistics we show in Fig.~\ref{fig_3d3q_h} the energy histogram 
and its error bars.
The histogram exhibits a {\bf double peak} structure, which is 
typically obtained when systems with first order transitions are 
simulated on finite lattices in the neighborhood of so called 
{\bf pseudo-transition temperatures}. These are finite lattice 
temperature definitions, which converge with increasing system size 
towards the infinite volume transition temperature. Equal heights 
of the maxima of the two peaks is one of the popular definition
of a pseudo-transition temperature for first order phase transitions.
Equal weights (areas under the curves) is another, used in the
lecture by Prof. Landau.
Our $\beta$ value needs to be re-weighted to a slightly higher 
value to arrange for equal heights (assignment {\tt a0303\_10}).
Our mean energy per spin, corresponding to the histogram of the figure 
is $e_s=-1.397\,(13)$. Due to the double peak structure of the 
histogram the error bar is relatively large. Still, the central 
limit theorem works and a Kolmogorov test shows that our statistics 
is large enough to create an approximately Gaussian distribution for 
the binned data (assignment {\tt a0303\_11}).

\begin{figure}[tb]
\vspace{5pc}
 \centerline{\hbox{ \psfig{figure=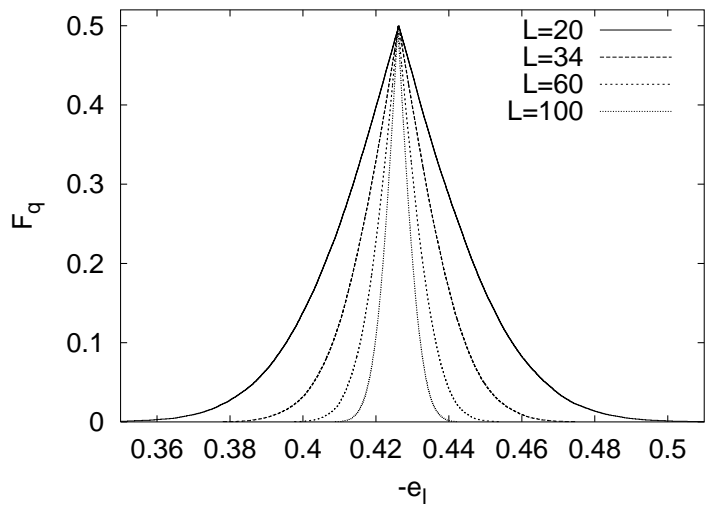,width=10cm} }}
\caption{Peaked distribution functions for the $O(3)$ $\sigma$-model 
mean energy per link on various lattices at $\beta=1.1$ 
(assignment {\tt a0304\_08}).} \label{fig_qdf_o3}
\end{figure}

{\bf Self-Averaging Illustration for the $O(3)$ model:}
\index{self-averaging}We compare in Fig.~\ref{fig_qdf_o3} the peaked 
distribution function of the mean energy per link $e_l$ for different 
lattice sizes. The property of {\bf self-averaging} is observed: The 
larger the lattice, the smaller the confidence range. The other way 
round, the peaked distribution function is very well suited to exhibit 
observables for which self-averaging does not work, as for instance 
encountered in spin glass simulations~\cite{BeBiJa00}.

 
\section{Statistical Errors of Markov Chain Monte Carlo 
         Data\label{AutoStat}} 

In large scale MC simulation it may take months, possibly years, to 
collect the necessary statistics. For such data a thorough error 
analysis is a must. A typical MC simulation falls into two parts: 

\begin{enumerate}

{\bf \item Equilibration:} Initial 
sweeps are performed to reach the equilibrium distribution. During 
these sweeps measurements are either not taken at all or they have 
to be discarded when calculating equilibrium expectation values.

{\bf \item Data Production:}
\index{data production}Sweeps with measurements are performed.
Equilibrium expectation values are calculated from this statistics.

\end{enumerate}

A rule of thumb is: {\bf Do not spend more than 50\% of your CPU time 
on measurements!} The reason for this rule is that one cannot be 
off by a factor worse than two ($\sqrt{2}$ in the statistical error).

How many sweeps should be discarded for reaching equilibrium?
In a few situations this question can be rigorously answered with 
the {\it Coupling from the Past} method (see the article by
W. Kendall in this volume). The next best thing to do is to measure 
the integrated autocorrelation time and to discard, after reaching 
a visually satisfactory situation, a number of sweeps which is 
larger than the integrated autocorrelation time. In practice even 
this can often not be achieved.

Therefore, it is re-assuring that it is sufficient to pick the number 
of discarded sweeps approximately right. With increasing statistics 
the contribution of the non-equilibrium data dies out like $1/N$, 
where $N$ is the number of measurements. This is eventually swallowed 
by the statistical error, which declines only like $1/\sqrt{N}$. The 
point of discarding the equilibrium configurations is that the factor 
in front of $1/N$ can be large.

There can be far more involved situations, like that the Markov 
chain ends up in a metastable configuration, which may even 
stay unnoticed (this tends to happen in complex systems like spin 
glasses or proteins).
 
\subsection{Autocorrelations} 

We like to estimate the expectation value $\widehat{f}$ of some physical
observable. We assume that the system has reached
equilibrium.  How many MC sweeps are needed to estimate
$\widehat{f}$ with some desired accuracy? To answer this question, one
has to understand the autocorrelations within the Markov chain.

Given is a {\bf time series} of $N$ measurements from a Markov process
\begin{equation} \label{time_series}
 f_i = f(x_i), ~~ i=1,\dots,N\ , 
\end{equation}
where $x_i$ are the configurations generated.
The label $i=1,\dots,N$ runs in the temporal 
order of the Markov chain and the elapsed time (measured in updates 
or sweeps) between subsequent measurements $f_i$, $f_{i+1}$ is always 
the same.  The estimator of the 
\index{expectation value}{expectation value} $\widehat{f}$ is
\begin{equation} \label{fhat_estimator}
 \overline{f} = {1\over N} \sum f_i\ .
\end{equation}
With the notation $$t=|i-j|$$ the definition of the 
\index{autocorrelations}{\bf autocorrelation function} 
of the observable $\widehat{f}$ is 
\begin{equation} \label{autocorrelation}
 \widehat{C}(t) = \widehat{C}_{ij} = \langle\, (f_i-\langle f_i\rangle)\, 
 (f_j - \langle f_j\rangle)\, \rangle =
 \langle f_i f_j\rangle - \langle f_i \rangle\, \langle f_j\rangle = 
 \langle f_0 f_t\rangle - \widehat{f}^{\,2} \end{equation}
where we used that translation invariance in time holds for the 
equilibrium ensemble. The asymptotic behavior for large $t$ is
\begin{equation} \label{tau_exp}
 \widehat{C}(t) \sim \exp \left( - {t\over \tau_{\rm exp}} \right)
 ~~{\rm for}~~ t \to \infty ,
\end{equation}
where $\tau_{\rm exp}$ is called\index{autocorrelations} 
{\bf (exponential) autocorrelation time} 
and is related to the second largest eigenvalue $\lambda_1$ of 
the transition matrix by $\tau_{\rm exp}=-\ln\lambda_1$
under the assumption that $f$ has a non-zero projection on the
corresponding eigenstate. Superselection rules are possible so that 
different autocorrelation times reign for different operators.

The variance of $f$ is a special case of the 
autocorrelations~(\ref{autocorrelation})
\begin{equation} \label{C0}
 \widehat{C} (0) = \sigma^2 (f)\ .
\end{equation}
Some algebra \cite{Berg} shows that the variance of the estimator 
$\overline{f}$~(\ref{fhat_estimator}) for the mean and the 
autocorrelation functions~(\ref{autocorrelation}) are related by
\begin{equation} \label{variance_correlated}
 \sigma^2(\overline{f})\ =\ {\sigma^2(f)\over N} \left[ 1 +
 2 \sum_{t=1}^{N-1} \left( 1 -{t\over N} \right)\, \widehat{c}(t) \right]
 ~~{\rm with}~~ \widehat{c}(t) = {\widehat{C}(t)\over \widehat{C}(0)}\ .
\end{equation}
This equation ought to be compared with the corresponding equation 
for uncorrelated random variables
$ \sigma^2(\overline{f}) = {\sigma^2(f)/ N}$. The difference is the 
factor in the bracket of~(\ref{variance_correlated}), which defines 
the\index{integrated autocorrelation time}
\index{autocorrelations}{\bf integrated autocorrelation time}
\begin{equation} \label{ia_time_N}
 \tau_{\rm int}\ =\ \left[ 1 + 2 \sum_{t=1}^{N-1} 
 \left(1-{t\over N}\right)\, \widehat{c}(t) \right]\ .
\end{equation}
For correlated data the variance of the mean is by the factor 
$\tau_{\rm int}$ larger than the corresponding\index{naive variance}
\index{variance}{\bf naive variance} for uncorrelated data:
\begin{equation} \label{variance_ratio}
  \tau_{\rm int}\ =\ {\sigma^2(\overline{f})\over 
  \sigma^2_{\rm naive} (\overline{f}) } ~~{\rm with}~~
  \sigma^2_{\rm naive} = {\sigma^2(f)\over N}\ .
\end{equation}
In most simulations one is interested in the limit $N\to\infty$ and 
equation~(\ref{ia_time_N}) becomes
\begin{equation} \label{ia_time}
\tau_{\rm int}\ =\ 1 + 2 \sum_{t=1}^{\infty} \widehat{c}(t)\ . 
\end{equation}
The numerical estimation of the integrated autocorrelation time
faces difficulties. Namely, the variance of the $N\to\infty$ 
estimator of $\tau_{\rm int}$ diverges:
\begin{equation}
\overline{\tau}_{\rm int}\ =\ 1 + 2 \sum_{t=1}^{\infty}
\overline{c}(t) ~~{\rm and}~~
\sigma^2 (\overline{\tau}_{\rm int})\ \to\ \infty\ ,
\end{equation}
because for large $t$ each $\overline{c}(t)$ adds a constant amount of 
noise, whereas the signal dies out like $\exp (-t/\tau_{\rm exp})$. 
To obtain an estimate one considers the $t$-dependent estimator
\begin{equation} \label{iat_estimate}
\overline{\tau}_{\rm int}(t)\ =\ 1 + 2 \sum_{t'=1}^t
\overline{c}(t') 
\end{equation}
and looks out for a {\bf window} in $t$ for which 
$\overline{\tau}_{\rm int}(t)$ is flat. 

To give a simple example, let us assume that the autocorrelation 
function is governed by a single exponential autocorrelation time
\begin{equation} \label{acor_simple}
  \widehat{C}(t)\ =\ {\rm const}\, 
  \exp \left( - {t\over\tau_{\rm exp}}\right)\ .
\end{equation}
In this case we can carry out the sum~(\ref{ia_time}) for the 
integrated autocorrelation function and find
\begin{equation} \label{iat_simple}
 \tau_{\rm int} = 1 + 2 \sum_{t=1}^{\infty}
 e^{-t/\tau_{\rm exp}} = 1 + { 2\,e^{-1/ \tau_{\rm exp}} 
 \over 1 - e^{-1/ \tau_{\rm exp} } }\ .
\end{equation}
For a large exponential autocorrelation time $\tau_{\rm exp}\gg 1$
the approximation
\begin{equation} \label{iat_simple_approx}
 \tau_{\rm int} = 1 + { 2\,e^{-1/ \tau_{\rm exp}} \over
 1 - e^{-1/ \tau_{\rm exp} } } \cong
 1 + { 2 - 2 / \tau_{\rm exp} \over 1 / \tau_{\rm exp} } =
 2\,\tau_{\rm exp} - 1\ \cong\ 2\,\tau_{\rm exp} 
\end{equation}
holds.

\subsection{Integrated Autocorrelation Time and Binning}

Using binning the integrated autocorrelation time can also be estimated 
via the \index{variance ratio}variance ratio. We bin the time 
series~(\ref{time_series}) into $N_{bs}\le N$ bins of
\begin{equation} \label{Nb}
 N_b = {\tt NBIN} = \left[ N\over N_{bs} \right] =
 \left[ {\tt NDAT} \over {\tt NBINS} \right]
\end{equation}
data each. Here $[.]$ stands for Fortran integer division, {\it i.e.}, 
$N_b={\tt NBIN}$ is the largest integer $\le N/N_{bs}$, implying
$N_{ba}\cdot N_b\le N$. It is convenient to choose the values of $N$ 
and $N_{bs}$ so that $N$ is a multiple of $N_{bs}$. The binned data 
are the averages
\begin{equation} \label{binned_data}
 f^{N_b}_j = {1\over N_b} \sum_{i=1+(j-1)N_b}^{jN_b} f_i 
~~~~{\rm for}~~ j=1,\dots ,N_{bs}\ .
\end{equation}
For $N_b > \tau_{\rm exp}$ the autocorrelations are essentially reduced 
to those between nearest neighbor bins and even these approach zero 
under further increase of the binsize.

For a set of $N_{bs}$ binned data $f^{N_b}_j$, $(j=1,\dots ,N_{bs})$ we 
may calculate the mean with its naive error bar. Assuming for the 
moment an infinite time series, we find the integrated autocorrelation 
time~(\ref{variance_ratio}) from the following ratio of sample 
variances
\begin{equation} \label{iat_binning} 
\tau_{\rm int}\ =\ \lim_{N_b\to\infty}\, \tau_{\rm int}^{N_b}\ 
~~{\rm with}~~\ \tau_{\rm int}^{N_b}\ =\ \left( 
{ s^2_{\overline{f}^{N_b}} \over s^2_{\overline{f}} }\right)\ .
\end{equation}
In practice the $N_b\to\infty$ limit will be reached for a sufficiently 
large, finite value of $N_b$. The statistical error of the 
$\tau_{\rm int}$ estimate~(\ref{iat_binning}) is, in the first 
approximation, determined by the errors of $s^2_{\overline{f}^{N_b}}$. 
The typical situation is then that, due to the central limit theorem, 
the binned data are approximately Gaussian, so that the {\bf error of 
$s^2_{\overline{f}^{N_b}}$ is analytically known} from the $\chi^2$ 
distribution. Finally, the fluctuations of 
$s^2_{\overline{f}}$ of the denominator give rise to a small
correction which can be worked out~\cite{Berg}. 

Numerically most accurate estimates of $\tau_{\rm int}$ are obtained 
for the finite binsize $N_b$ which is just large enough that the binned 
data~(\ref{binned_data}) are practically uncorrelated. 
While the Student distribution shows that the confidence intervals 
of the error bars from 16 uncorrelated normal data are reasonable 
approximations to those of the Gaussian standard deviation,
about 1000 independent data are needed to provide a decent 
estimate of the corresponding variance (at the 95\% confidence 
level with an accuracy of slightly better than 10\%). 
It makes sense to work with error bars from 16 binned data, but the 
error of the error bar, and hence a reliable estimate of 
$\tau_{\rm int}$, requires far more data.
 
\subsection{Illustration: Metropolis generation of normally
             distributed data} 

We generate normally distributed data according to the Markov process
\begin{equation} \label{gau_metro}
\index{Metropolis algorithm}
 x' = x + 2\,a\,x^r - a
\end{equation}
where $x$ is the event at hand, $x^r$ a uniformly distributed 
random number in the range $[0,1)$, and the real number $a>0$ is
a parameter which relates to the efficiency of the algorithm. The 
new event $x'$ is accepted with the Metropolis probability
\begin{equation} \label{mg_accept}
 P_{\rm accept}(x') = \begin{cases} 
                       1~~{\rm for}~~x'^{\,2}\le x^2;\\
  \exp [-( x'^{\,2}-x^2)/2 ]\ {\rm for}\ x'^{\,2}>x^2 .
  \end{cases} 
\end{equation}
If $x'$ is rejected, the event $x$ is counted again. The Metropolis
process introduces an autocorrelation time in the generation of 
normally distributed random data.

\begin{figure}[tb]
\vspace{5pc}
 \centerline{\hbox{ \psfig{figure=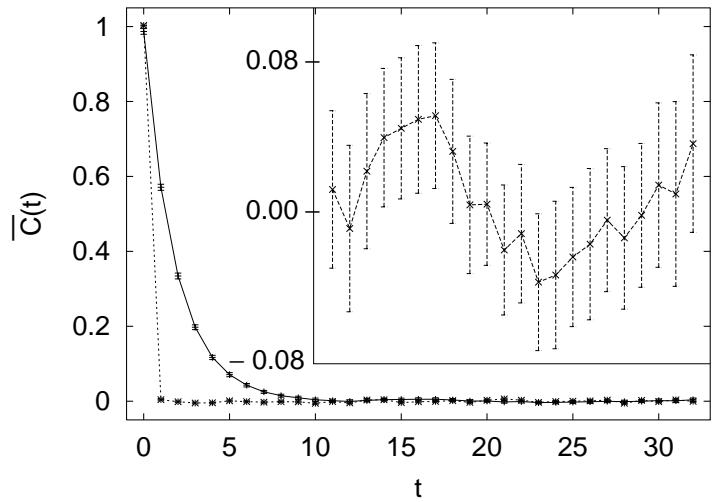,width=10cm} }}
\caption{The autocorrelation function~(\ref{autocorrelation}) of a 
Metropolis time series for the normal distribution (upper data) in 
comparison with those of our Gaussian random number generator (lower 
data). For $t\ge 11$ the inlay shows the autocorrelations on an 
enlarged ordinate. The straight lines between the data points are 
just to guide the eyes. The curves start with 
$\overline{C}(0)\approx 1$ because the variance 
of the normal distribution is one.} \label{fig_acor}
\end{figure}

We work with $N=2^{17}=131072$ data and take $a=3$ 
for the Markov process~(\ref{gau_metro}), what gives an acceptance rate 
of approximately 50\%. The autocorrelation function of this process
is depicted in Fig.~\ref{fig_acor} (assignment {\tt a0401\_01}).
\begin{figure}[tb]
\vspace{5pc}
 \centerline{\hbox{ \psfig{figure=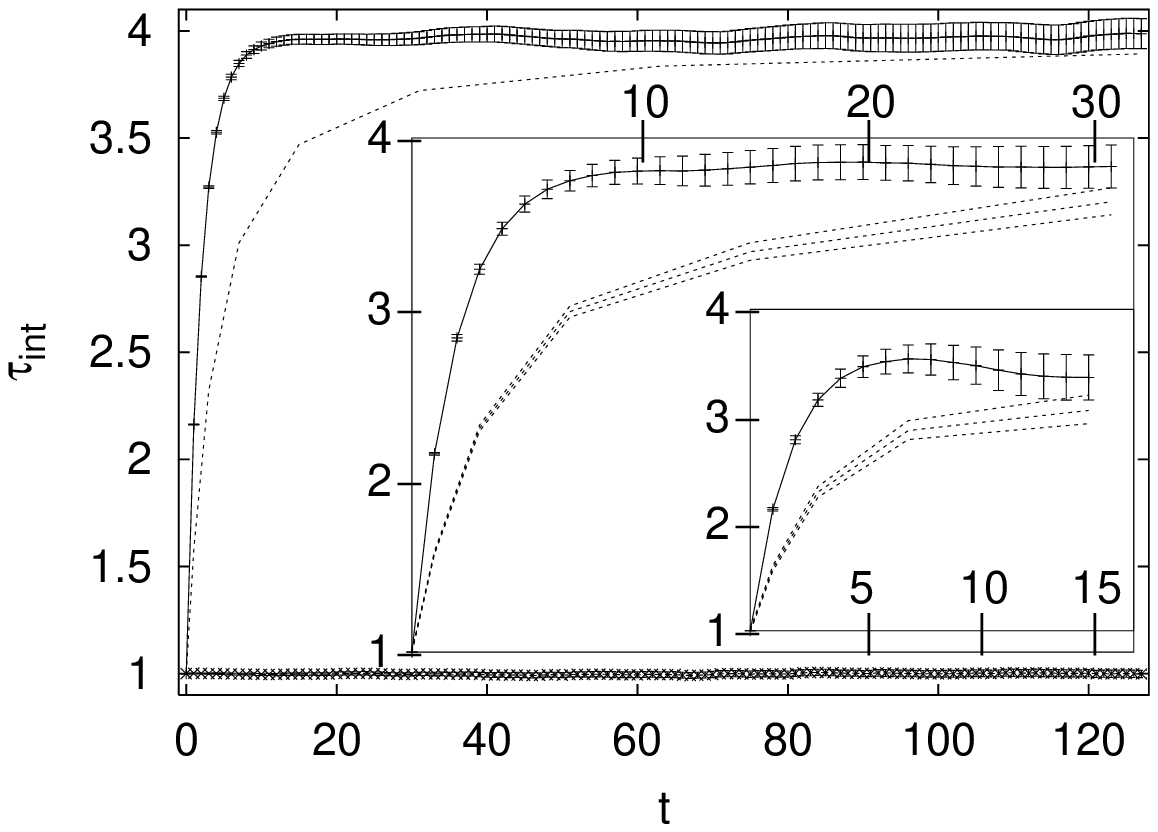,width=10cm} }}
\caption{The upper curves in the figure and its inlays display the 
estimators obtained by direct calculation. The lowest curve is for 
the Gaussian random number generator. The remaining curves are binning 
procedure estimators of the integrated autocorrelation time with one 
standard deviation bounds. The main figure relies on $2^{21}$ data and
depicts estimators up to $t=127$. The first inlay relies on $2^{17}$
data and depicts estimators up to $t=31$. The second inlay 
relies on $2^{14}$ data and depicts estimators up to $t=15$.}
 \label{fig_at_int}
\end{figure}
The integrated autocorrelation time (assignment {\tt a0401\_02}) is
shown in Fig.~\ref{fig_at_int}.  We compare the $\tau^{N_b}_{\rm int}$ 
estimators with the direct estimators $\tau_{\rm int}(t)$ at
\begin{equation} \label{t_eq_Nbm1}
 t = N_b -1\ .
\end{equation}
With this relation the estimators agree for binsize $N_b=1$ and for 
larger $N_b$ the relation gives the range over which we combine data 
into either one of the estimators. The approach of the binning procedure 
towards the asymptotic $\tau_{\rm int}$ value is slower than that of the 
direct estimate of $\tau_{\rm int}$.

For our large ${\tt NDAT}=2^{21}$ data set $\tau_{int}(t)$ reaches its 
plateau before $t=20$. All the error bars within the plateau are 
strongly correlated. Therefore, it is not recommendable to make an 
attempt to combine them. Instead, it is save to pick an appropriate 
single value and its error bar as the final estimate:
\begin{equation} \label{iat_Gauss_estimate}
  \tau_{\rm int} = \tau_{\rm int} (20) = 3.962 \pm 0.024 ~~{\rm from}~~
  2^{21}=2,097,152\ {\rm data}.
\end{equation}
The binning procedure, on the other hand, shows an increase of 
$\tau^{N_b}_{\rm int}$ all the way to $N_b=2^7=128$, where the estimate 
with the one confidence level error bounds is
\begin{equation} \label{F_cl_21}
 3.85 \le \tau^{128}_{\rm int} \le 3.94 ~~{\rm from}~~
2^{14}=16,384\ {\rm bins}~~{\rm from}~~2^{21}\ {\rm data}.
\end{equation}

How many data are needed to allow 
for a meaningful estimate of the integrated autocorrelation time?

For a statistics of ${\tt NDAT}=2^{17}$ the autocorrelation signal 
disappears for $t\ge 11$ into the statistical noise.  Still, there is 
clear evidence of the hoped for window of almost constant estimates. 
A conservative choice is to take $t=20$ again, which now gives
\begin{equation} 
\tau_{\rm int} =
\tau_{\rm int}(20)=3.86\pm 0.11 ~~{\rm from}~~ 2^{17}\ {\rm data}.
\end{equation}
Worse is the binning estimate, which for the $2^{17}$  data is 
\begin{equation} \label{F_cl_17}
 3.55 \le \tau^{32}_{\rm int} \le 3.71 ~~{\rm from}~~
2^{12}=4,096\ {\rm bins}~~{\rm from}~~2^{17}=131,072\ {\rm data}.
\end{equation}
Our best value~(\ref{iat_Gauss_estimate}) is no longer covered by 
the two standard deviation zone.

For the second inlay the statistics is reduced to ${\tt NDAT}=2^{14}$. 
With the integrated autocorrelation time rounded to 4, this 
is 4096 times $\tau_{\rm int}$. For binsize $N_b=2^4=16$ we are then 
down to $N_{bs}=1024$ bins, which are needed for accurate error bars 
of the variance. To work with this 
number we limit, in accordance with equation~(\ref{t_eq_Nbm1}), our 
$\tau_{\rm int}(t)$ plot to the range $t\le 15$. Still, we find a 
quite nice window of nearly constant $\tau_{\rm int}(t)$, namely 
all the way from $t=4$ to $t=15$. By a statistical fluctuation
(assignment {\tt a0401\_03}) $\tau_{\rm int}(t)$ takes its maximum 
value at $t=7$ and this makes $\tau_{\rm int}(7) = 3.54\pm 0.13$ a 
natural candidate. However, this value is inconsistent with our best 
estimate~(\ref{iat_Gauss_estimate}).
The true $\tau_{\rm int}(t)$ increases monotonically as function of 
$t$, so we know that the estimators have become bad for $t>7$. 
The error bar at $t=7$ is too small to take care of our
difficulties. One may combine the $t=15$ error bar with the $t=7$
estimate. In this way the result is 
\begin{equation}  \label{tau_int14}
\tau_{\rm int}=3.54\pm 0.21~~{\rm for}~~2^{14}=16,384\ {\rm data},
\end{equation}
which achieves consistency with~(\ref{iat_Gauss_estimate}) in the 
two error bar range. For binsize $N_b=16$ the binning estimate is 
\begin{equation} \label{F_cl_example}
 2.93 \le \tau^{16}_{\rm int} \le 3.20 ~~{\rm from}~~
 2^{10}=1,024\ {\rm bins}~~{\rm from}~~2^{14}\ {\rm data}. 
\end{equation}
Clearly, the binsize $N_b=16$ is too small for an estimate of the 
integrated autocorrelation time.  We learn that 
one needs a binsize of at least ten times the integrated autocorrelation 
time $\tau_{\rm int}$, whereas for its direct estimate it is sufficient 
to have $t$ about four times larger than $\tau_{\rm int}$.
 
\section{Self-consistent versus reasonable error analysis}

By visual inspection of the time series, one may get an impression about 
the length of the out-of-equilibrium part of the simulation. On top of 
this one should still choose
\begin{equation} \label{auto_equi}
{\tt nequi}\ \gg\ \tau_{\rm int}\ ,
\end{equation}
to allow the system to settle. That is a first reason, why it appears 
necessary to control the integrated autocorrelation time of a MC 
simulations. A second reason is that we have to control the error 
bars of the equilibrium part of our simulation. Ideally the error 
bars are calculated as
\begin{equation} \label{error_best}
\triangle \overline{f}\ =\ \sqrt{\sigma^2(\overline{f})}
~~~{\rm with}~~~
\sigma^2(\overline{f})\ =\ \tau_{\rm int}\,{\sigma^2(f)\over N}\ .
\end{equation}
This constitutes a {\bf self-consistent error analysis} of a MC 
simulation.

However, the calculation of the integrated autocorrelation time may
be out of reach.  Many more than the about twenty independent data are 
needed, which according to the Student distribution are sufficient to 
estimate mean values with reasonably reliable error bars.

In practice, one has to be content with what can be done. {\bf Often 
this means to rely on the binning method.} We simply calculate error 
bars of our ever increasing statistics with respect to a fixed number of 
\begin{equation} \label{NBINS_choice}
{\tt NBINS}\ \ge\ 16\ .
\end{equation}
In addition, we may put 10\% of the 
initially planned simulation time away for reaching equilibrium. 
{\it A-posteriori}, this can always be increased. Once the statistics 
is large enough, our small number of binned data become effectively 
independent and our error analysis is justified. 

How do we know that the statistics has become large enough? In practical 
applications there can be indirect arguments, like FSS estimates, which 
tell us that the integrated autocorrelation time is in fact (much) 
smaller than the achieved bin length. This is no longer self-consistent, 
as we perform no explicit measurement of $\tau_{\rm int}$, but it is a 
\index{error bar}{\bf reasonable error analysis}.

\section{Comparison of Markov chain MC algorithms}

\begin{figure}[tb]
\vspace{5pc}
 \centerline{\hbox{ \psfig{figure=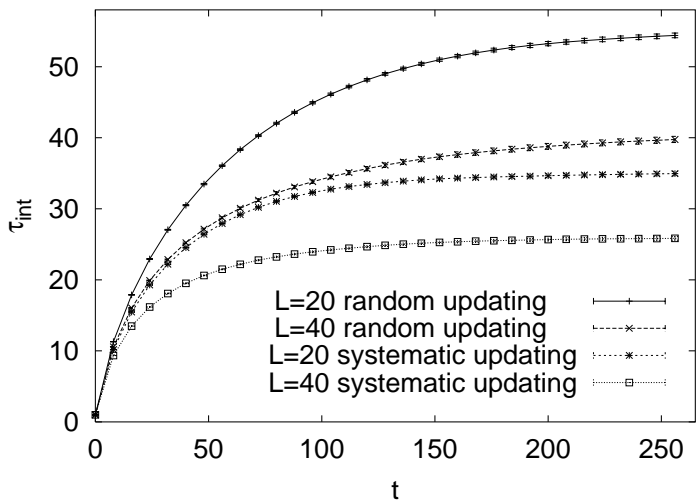,width=10cm} }}
\caption{Comparison of the integrated autocorrelation time of the 
Metropolis process with random updating versus sequential updating 
for the $d=2$ Ising model at $\beta =0.4$ (assignment 
{\tt a0402\_01} B). The ordering of the curves is 
identical with the ordering of the labels in the figure. }
 \label{fig_iat1r}
\end{figure}

Is the 1-hit Metropolis algorithm more efficient with sequential 
updating or with random updating? For $2d$ Ising lattices at 
$\beta =0.4$ Fig.~\ref{fig_iat1r} illustrates that sequential
updating wins. This is apparently related to the fact that 
random updating may miss out on some spins for some time, whereas
sequential updating touches each spin with certainty during one
sweep.

\begin{figure}[tb]
\vspace{5pc}
 \centerline{\hbox{ \psfig{figure=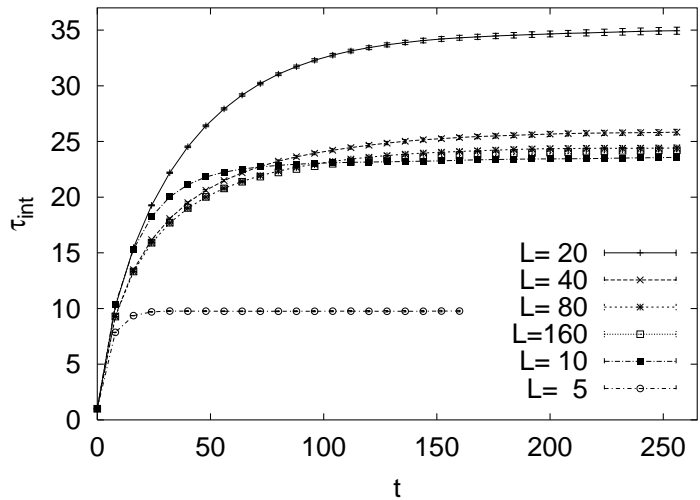,width=10cm} }}
\caption{One-hit Metropolis algorithm with sequential updating: Lattice 
size dependence of the integrated autocorrelation time for the $d=2$ 
Ising model at $\beta =0.4$ (assignment {\tt a0402\_01}~A). The 
ordering of the curves is identical with the ordering of the labels 
in the figure. }
 \label{fig_iat1}
\end{figure}

\begin{figure}[tb]
\vspace{5pc}
 \centerline{\hbox{ \psfig{figure=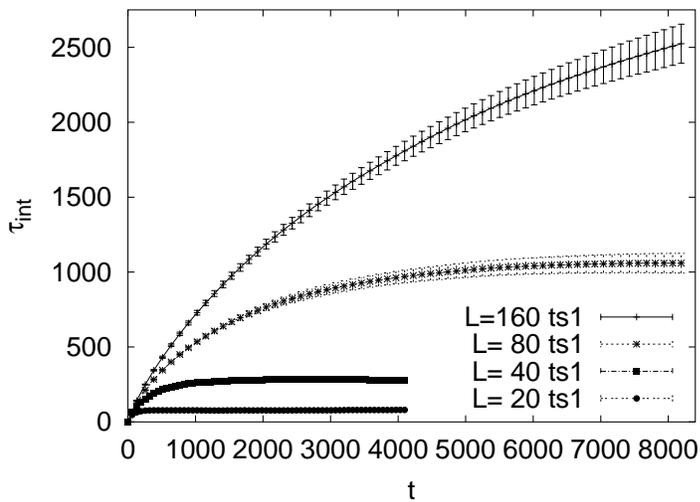,width=10cm} }}
\caption{One-hit Metropolis algorithm 
with sequential updating: Lattice size dependence of the integrated 
autocorrelation time for the $d=2$ Ising model at its critical 
temperature (assignment {\tt a0402\_02}~D). The ordering of the 
curves is identical with the ordering of the labels in the figure.}
 \label{fig_iat2_large}
\end{figure}

Figures~\ref{fig_iat1} and~\ref{fig_iat2_large} illustrate $2d$ Ising 
model simulations off and on the critical point. Off the critical point, 
at $\beta =0.4$, the integrated autocorrelation time increases for
$L=5,\,10$ and~20. Subsequently, it decreases to approach for 
$L\to\infty$ a finite asymptotic value. On the critical point,
at $\beta=\beta_c=\ln(1+\sqrt{2})/2$, 
\index{critical slowing down}{\bf critical slowing down} is
observed, an increase $\tau_{\rm int}\sim L^z$ with lattice
size, where $z\approx 2.17$ is the 
\index{dynamics}{\bf dynamical critical exponent}. 
of the $2d$ Ising model. Estimates of $z$ are compiled in the book 
by Landau and Binder~\cite{LaBiBook}.

\begin{figure}[tb]
\vspace{5pc}
 \centerline{\hbox{ \psfig{figure=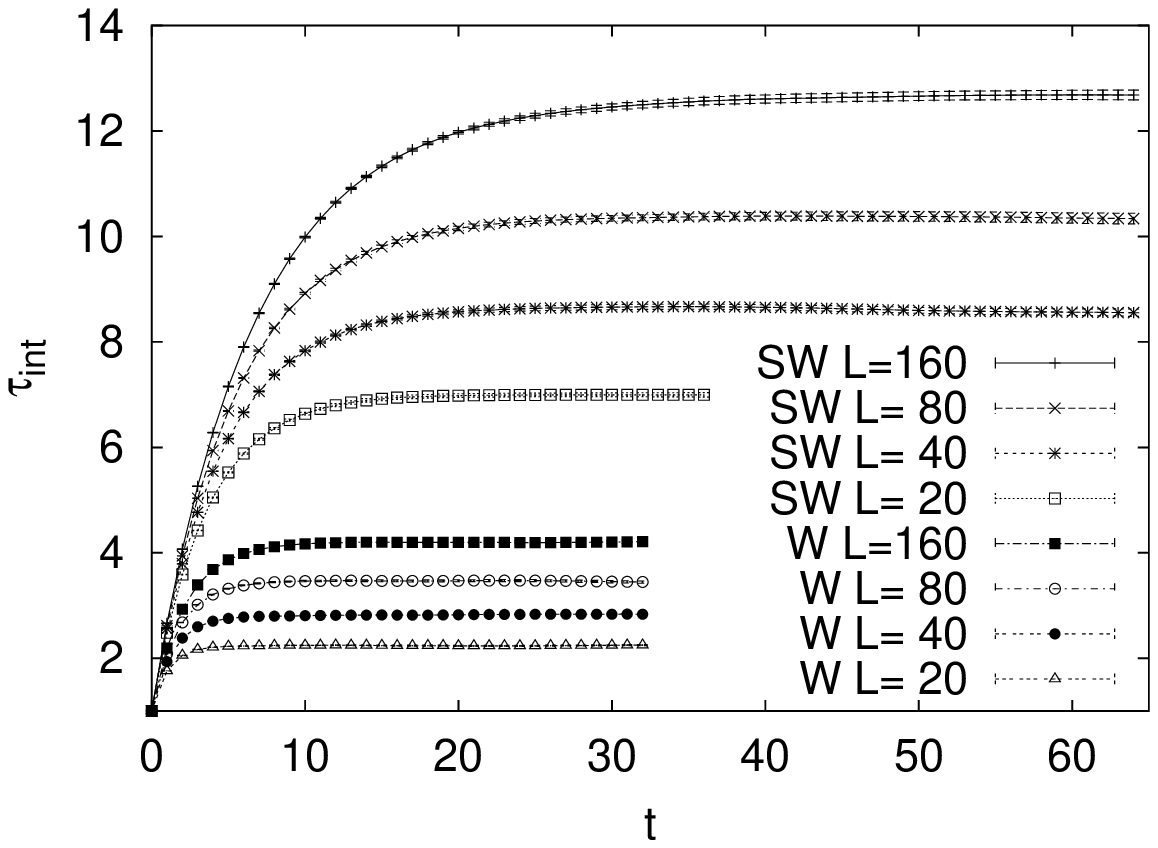,width=10cm} }}
\caption{Estimates of integrated autocorrelation times from 
simulations of the $d=2$ Ising model at the critical temperature 
$\beta_c=0.44068679351$ (assignment {\tt a0503\_05}). }
 \label{fig_at_cl}
\end{figure}

Using another MC dynamics the critical slowing down can be overcome.
Fig.~\ref{fig_at_cl} shows the major improvements for Swendsen-Wang 
\cite{SwWa87} (SW) and Wolff \cite{Wo89} (W) cluster updating.

\begin{figure}[tb]
\vspace{5pc}
 \centerline{\hbox{ \psfig{figure=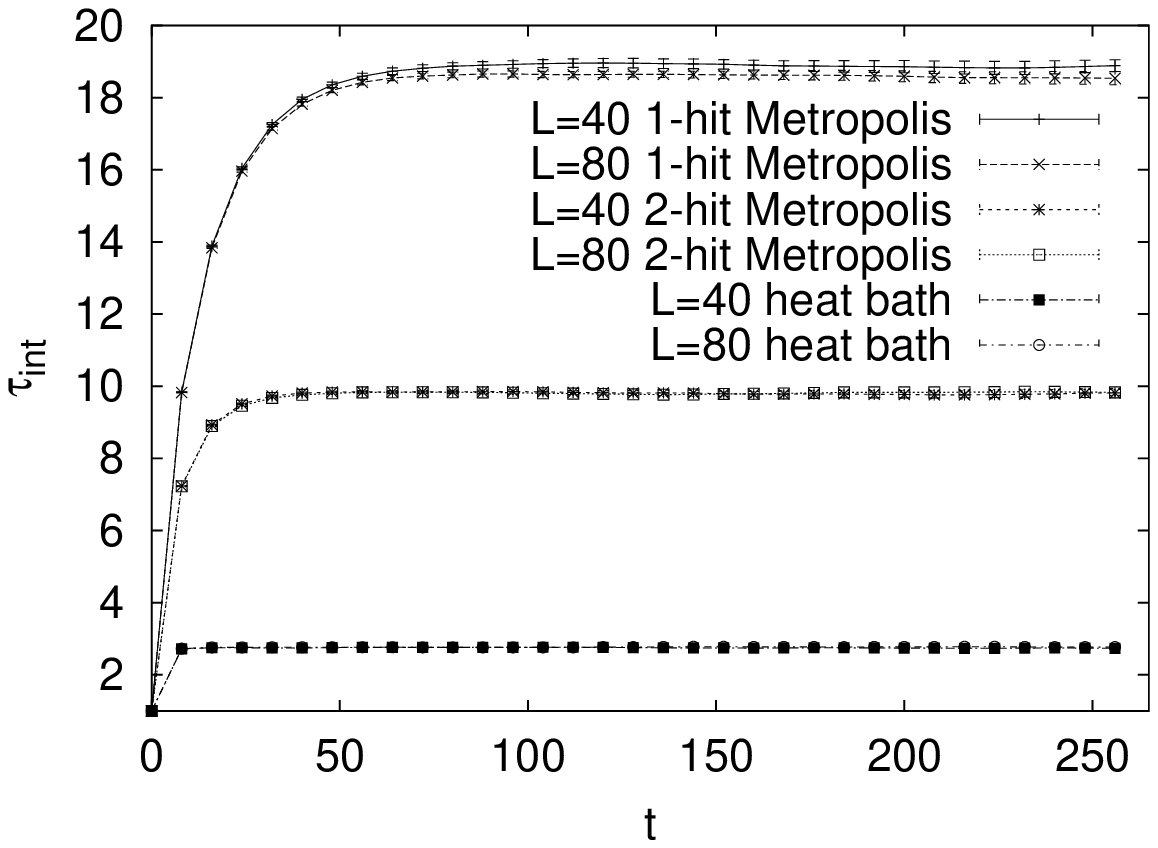,width=10cm} }}
\caption{Systematic updating: Comparison of the integrated 
autocorrelation times of the 1-hit and 2-hit Metropolis algorithms 
and the heat bath algorithm for the 10-state Potts model on $L\times L$ 
lattices at $\beta =0.62$ (assignment {\tt a0402\_06}). The $L=40$ 
and $L=80$ curves lie almost on top of one another. }
 \label{fig_iathb}
\end{figure}

Finally, Fig.~\ref{fig_iathb} exhibit the improvements of heat bath 
over Metropolis updating for the 10-state $d=2$ Potts model at 
$\beta=0.62$.


\section{Multicanonical Simulations}
 
One of the questions which ought to be addressed before performing
a large scale computer simulation is {``What are suitable weight
factors for the problem at hand?''} So far we used the Boltzmann 
weights as this appears natural for simulating the canonical ensemble. 
However, a broader view of the issue is appropriate.  

Conventional, canonical simulations calculate expectation values at 
a fixed temperature $T$ and can, by re-weighting techniques, only 
be extrapolated to a vicinity of this temperature. For 
multicanonical simulations this is different. A single simulation 
allows to obtain equilibrium properties of the Gibbs ensemble
over a range of temperatures. Of particular interest are two 
situations for which canonical simulations do not provide the 
appropriate implementation of importance sampling: 
\begin{enumerate}
\item The physically important configurations are rare in the 
      canonical ensemble.
\item A rugged free energy landscape makes the physically important 
      configurations difficult to reach. 
\end{enumerate}

MC calculation of the interface tension of a first order phase 
transition provide an example for the first situation. Let $N=L^d$ be 
the lattice size. For first order phase transition {\bf 
pseudo-transition temperatures} $\beta^c (L)$ exist so that the energy 
distributions $P(E)=P(E;L)$ become double peaked and the maxima at 
$E^1_{\max} < E^2_{\max}$ 
are of equal height $P_{\max} = P (E^1_{\max}) = P (E^2_{\max})$.
In-between the maximum values a minimum is located at some energy 
$E_{\min}$.  Configurations at $E_{\min}$ are exponentially 
suppressed like
\begin{equation} \label{Pmin}
P_{\min} = P (E_{\min}) = c_f\, L^p\, \exp ( - f^s A )
\end{equation}
where $f^s$ is the interface tension and $A$ is the minimal area 
between the phases, $A=2L^{d-1}$ for an $L^d$ lattice, $c_f$ and
$p$ are constants (computations of $p$ have been done in the 
capillary-wave approximation). The interface tension can be determined 
by \index{Binder histogram method}{Binder's histogram method~\cite{Bi82}. 
One has to calculate the quantities
\begin{equation} \label{Itension}
f^s (L) = - {1\over A(L)}\, \ln R(L)\ ~~~~{\rm with}~~~~\ 
R(L) = {P_{\min} (L) \over P_{\max} (L) }
\end{equation}
and to make a FSS extrapolation of $f^s(L)$ for $L\to\infty$. 

For large systems a canonical MC
simulation will practically never visit configurations at energy 
$E=E_{\min}$ and estimates of the ratio $R(L)$ will be very inaccurate. 
The terminology \index{supercritical slowing down}{\bf supercritical 
slowing down} was coined to characterize such an exponential 
deterioration of simulation results with lattice size.

\index{multicanonical}{\bf Multicanonical simulations} \cite{BeNe92}  
approach this problem by sampling, in an appropriate energy range, 
with an {\bf approximation} to the weights 
\begin{equation} \label{wspectral}
 \index{multicanonical!weights}\index{weights}
 w_{1/n} (E^{(k)}) = {1\over n(E^{(k)})} = 
 \exp \left[ -b\left(E^{(k)}\right)\,E + a\left(E^{(k)}\right) \right]
\end{equation}
where $n(E)$ is the number of states of energy $E$. The function $b(E)$ 
defines the inverse \textbf{microcanonical temperature} and $a(E)$ the 
{\bf dimensionless, microcanonical free energy}. The function $b (E)$ 
has a relatively smooth dependence on its arguments, which makes it a 
useful quantity when dealing with the weight factors.  

Instead of the canonical energy distribution $P(E)$,
one samples a new multicanonical distribution
\begin{equation} \label{Pmuca}
 P_{mu} (E) = c_{mu}\, n(E)\, w_{mu} (E) \approx c_{mu}\ .
\end{equation}
The desired canonical probability density is obtained by
re-weighting
\begin{equation} \label{reweight}
 P (E) = {c_{\beta}\over c_{mu}}\, {P_{mu}(E) \over w_{mu} (E)}\,
         e^{-\beta E} .
\end{equation}
This relation is rigorous, because the weights $w_{mu}(E)$ used in
the simulation are exactly known. Accurate estimates of the
interface tension (\ref{Itension}) become possible.

The multicanonical method requires {two steps:}

\begin{enumerate}

\item Obtain a {working estimate} $\widehat{w}_{mu}(k)$ of the weights 
      $\widehat{w}_{1/n}(k)$. Working estimate means that the 
      approximation to (\ref{wspectral}) has to be good enough 
      to ensure movement in the desired energy range.

\item Perform a Markov chain MC simulation with the {fixed} weights 
      $\widehat{w}_{mu}(k)$. The thus generated configurations
      constitute the {multicanonical ensemble}. Canonical expectation 
      values are found by re-weighting to the Gibbs ensemble and 
      jackknife methods allow reliable error estimates.

\end{enumerate}

It is a strength of computer simulations that one can generate
artificial (not realized by nature) ensembles, which enhance the 
probabilities of rare events one may be interested in, or speed up 
the dynamics.
Nowadays {Generalized Ensembles} (umbrella, multicanonical, 1/k, 
...) have found many applications. Besides for first order phase 
transitions they are in particular usefull for complex systems 
such as biomolecules, where they accelerate the dynamics. For a 
review see~\cite{HaOk99}.

\subsection{How to get the Weights?}

To get the weights is at the heart of the method. Some approaches are:

\begin{enumerate}
\item Overlapping, constrained (microcanonical) MC simulations. A
      potential problem is to fulfill ergodicity.
\item FSS Estimates. This appears to be best when it works, but
      there may be no FSS theory.
\item General Purpose Recursions. Problem: They tend to deteriorate 
      with increasing lattice size (large lattices). 
\end{enumerate}

{\bf The Multicanonical Recursion} (a variant of \cite{Be96}){\bf :}
The multicanonical parameterization of the weights is
$$ w(a) = e^{-S(E_a)} = e^{-b(E_a)\, E_a + a(E_a)}\ , $$
{where (for {$\epsilon$} being the smallest energy stepsize)}
$$ b(E) = \left[ S(E+\epsilon) - S(E) \right] / \epsilon 
   ~~{\rm and}~~ a(E-\epsilon) =  a(E) 
  + \left[ b(E-\epsilon)-b(E) \right]\, E\ . $$
The recursion reads then (see \cite{Be03} for details):
$$ {b^{n+1}(E) = b^n (E) + \hat{g}^n_0(E)\,
 [ \ln H^n(E+\epsilon)-\ln H^n(E)] / \epsilon} $$
$$ \hat{g}^n_0 (E) = g^n_0(E)\, /\, [g^n(E) + \hat{g}^n_0 (E)]\, ,$$
$$ g^n_0 (E) = H^n (E+\epsilon)\, H^n (E)\, /\,
   [H^n (E+\epsilon) + H^n (E)]\, ,$$
$$ g^{n+1} (E) = g^n(E) + g^n_0(E),\ g^0(E)=0\, .$$

{\bf The Wang-Landau Recursion \cite{WaLa01}:}
Updates are performed with estimators $g(E)$ of the density of states
$$ p(E_1\to E_2) = \min \left[ {g(E_1)\over g(E_2)}, 1\right]\ .$$
Each time an energy level is visited, the estimator of $g(E)$ is  
updated according to 
$$ g(E) \to g(E)\,f $$
where, initially, $g(E)=1$ and $f=f_0=e^1$. Once the
desired energy range is covered, the factor $f$ is refined:
$$ f_1=\sqrt{f},\ f_{n+1}=\sqrt{f_{n+1}} $$
until some value very close to one like $f=1.00000001$ is reached. 
Afterwards the usual multicanonical production runs may be carried out.

\section{Multicanonical Example Runs ($2d$ Ising and Potts models)} 

Most illustrations of this section are from Ref.~\cite{Be03}.

For an Ising model on a $20\times 20$ lattice the multicanonical 
recursion is run in the range
\begin{equation} \label{2dI_namin_max}
{\tt namin} = 400 \le {\tt iact} \le 800 = {\tt namax}\ . 
\end{equation}
The recursion is terminated after a number of so called 
\index{tunneling} tunneling events. A {\bf tunneling event} is defined 
as an updating process which finds its way from
\begin{equation} \label{muca_tunneling}
{\tt iact} = {\tt namin} ~~~{\rm to}~~~ {\tt iact} = {\tt namax}
~~~{\rm and}~~{\rm back}\ . 
\end{equation}
This notation comes from applications to first order phase transitions. 
An alternative notation for tunneling event is {\bf random walk cycle}.  
For most applications 10 tunneling events lead to acceptable weights.

For the Ising model example run we find the requested 10 tunneling 
events after 787 recursions and 64,138 sweeps (assignment 
{\tt a0501\_01}). In assignment {\tt a0501\_02} a similar example 
run is performed for the $2d$ 10-state Potts model.

{\bf Performance:}
If the multicanonical weighting would remove all relevant {free energy 
barriers}, the behavior of the updating process would become that 
of a free {\bf random walk}.  Therefore, the theoretically optimal 
performance for the second part of the multicanonical simulation is
\begin{equation} \label{muca_RW}
{\tau_{\rm tun}\ \sim\ V^2\ .}
\end{equation}
Recent work about first order transitions by Neuhaus and Hager 
\cite{NeHa02} shows that the multicanonical procedure removes only 
the leading free energy barrier, while at least one subleading barrier 
causes a residual supercritical slowing done. Up to certain medium 
sized lattices the behavior {$V^{2+\epsilon}$} gives a rather good 
effective description. For large lattices exponential slowing down 
dominates again. The slowing down of the weight recursion with the 
volume size is expected to be even (slightly) worse than that of the 
second part of the simulation. 

{\bf Re-Weighting to the Canonical Ensemble:}
Let us assume that we have performed a multicanonical simulation
which covers the energy histograms for a temperature range
\begin{equation} \label{beta_range}
\beta_{\min}\ \le\ \beta = {1\over T}\ \le\ \beta_{\max}\ .
\end{equation}
Given the {multicanonical time series}, where $i=1,\dots,n$ labels 
the generated configurations, the formula 
\begin{equation} \label{O_muca}
 \index{multicanonical!canonical estimator} 
 \overline{\mathcal{O}} = { \sum_{i=1}^n \mathcal{O}^{(i)}\, \exp 
 \left[ -\beta\,E^{(i)} + b(E^{(i)})\,E^{(i)}-a(E^{(i)}) \right] \over 
 \sum_{i=1}^n \exp 
 \left[ -\beta\,E^{(i)} + b(E^{(i)})\,E^{(i)}-a(E^{(i)}) \right] }\ .
\end{equation}
replaces the multicanonical weighting of the simulation by the Boltzmann 
factor. The denominator differs from the partition function $Z$ by a 
constant factor which drops out.

For discrete systems it is sufficient to keep histograms
when only functions of the energy are calculated. For an operator 
$\mathcal{O}^{(i)} = f(E^{(i)})$ equation~(\ref{O_muca}) simplifies to
\begin{equation} \label{f_muca}
\overline{f} = { \sum_E f(E)\, h_{mu}(E)\, \exp 
\left[ -\beta\,E + b(E)\,E-a(E) \right] \over \sum_E h_{mu}(E)\, 
\exp \left[ -\beta\,E + b(E)\,E-a(E) \right] }
\end{equation}
where $h_{mu}(E)$ is the histogram sampled during the multicanonical
production run and the sums are over all energy values for which
$h_{mu}(E)$ has entries.

The computer implementation of these equations requires care. The 
differences between the largest and the smallest numbers encountered 
in the exponents can be really large. We can {avoid large numbers 
by dealing only with logarithms} of sums and partial sums. 
For $C=A+B$ with $A>0$ and $B>0$ we can calculate $\ln C = \ln (A+B)$ 
from the values $\ln A$ and $\ln B$, without ever storing either $A$ 
or $B$ or $C$ (see \cite{Berg} for more details):
\begin{eqnarray} \label{lnC}
\ln C &=& \ln \left[ \max (A,B)\,\left( 1+{\min (A,B)\over \max (A,B)}
\right) \right] \\ \nonumber
&=& \max \left( \ln A,\ln B \right) + \ln \{ 1 +
\exp \left[ \min (\ln A,\ln B) - \max (\ln A,\ln B) \right] \} 
\\ \nonumber &=& \max \left( \ln A,\ln B \right) + 
\ln \{ 1 + \exp \left[ - |\ln A - \ln B| \right] \}\ .
\end{eqnarray}

\subsection{Energy and Specific Heat Calculation}

\begin{figure}[tb] \vspace{5pc}
 \centerline{\hbox{ \psfig{figure=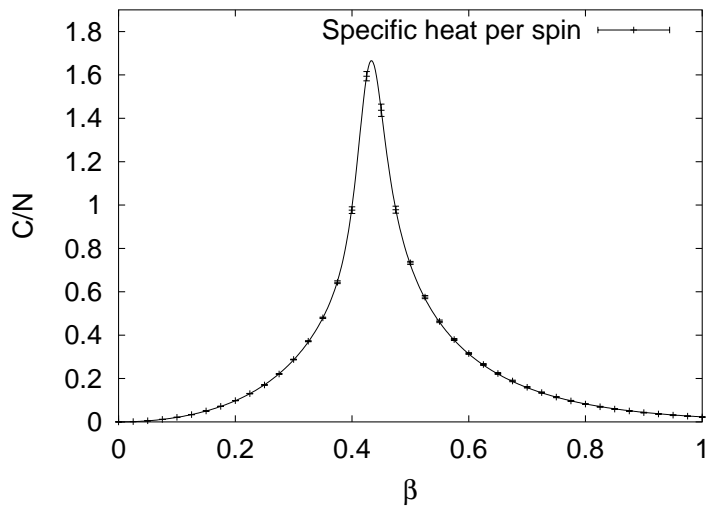,width=10cm} }}
 \caption{Specific heat per spin for the Ising model on a $20\times 20$
 lattice: Multicanonical data versus exact results of Ferdinand and 
 Fisher. This figure was first published in [6].} \label{fig_2dI_C}
\end{figure}

We are now ready to produce multicanonical data for the energy per spin 
of the $2d$ Ising model on a $20\times 20$ lattice (assignment 
{\tt a0501\_03}). The same numerical data allow to calculate the 
\index{specific heat}{\bf specific heat} defined by 
\begin{equation} \label{specific_heat}
  C\ =\ {d\,\widehat{E}\over d\,T}\ =\ \beta^2\, \left( \,
\langle E^2 \rangle - \langle E \rangle^2 \right)\ .
\end{equation}
The comparison of the multicanonical specific heat data with the
exact curve of Ferdinand and Fisher \cite{FeFi69} is shown in
Fig.~\ref{fig_2dI_C} (error bars rely on the jackknife method).

\begin{figure}[tb] \vspace{5pc}
 \centerline{\hbox{ \psfig{figure=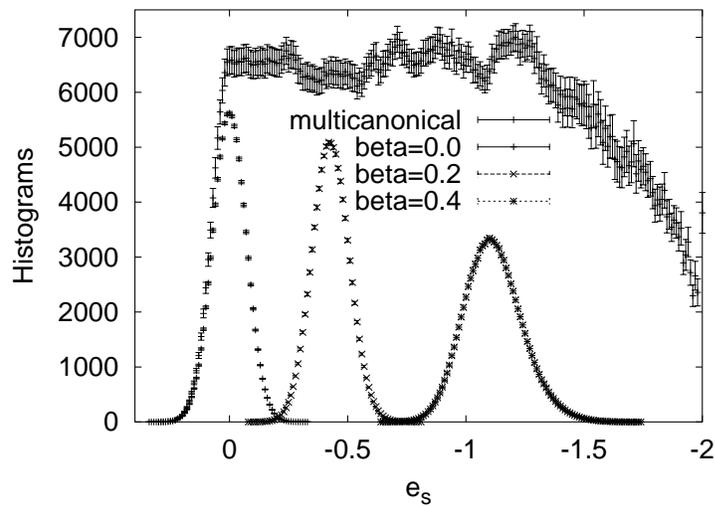,width=10cm} }}
 \caption{Energy histogram from a multicanonical simulation of the 
 Ising model on a $20\times 20$ lattice together with canonically
 re-weighted histograms (assignment {\tt a0501\_04}). This figure
 was first published in [6].} \label{fig_2dI_muh} 
\end{figure}

The energy histogram of this multicanonical simulation together its 
canonically re-weighted descendants at $\beta = 0$, $\beta = 0.2$ 
and $\beta =0.4$ is shown in Fig.~\ref{fig_2dI_muh}. 
The normalization of the multicanonical histogram is adjusted so that
it fits into the same figure with the three re-weighted histograms.

\begin{figure}[tb] \vspace{5pc}
 \centerline{\hbox{\psfig{figure=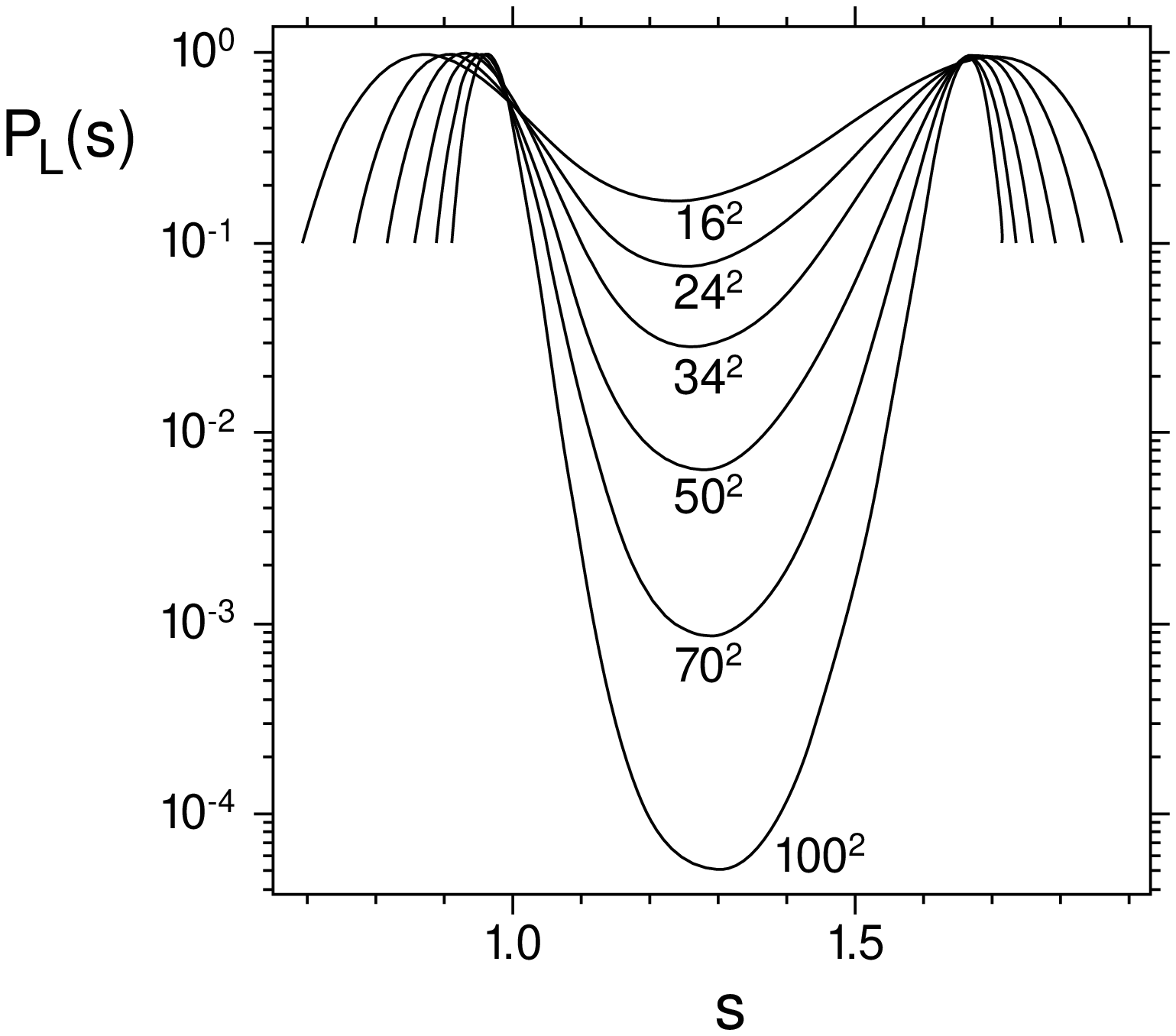,width=10cm}}}
 \caption{Energy histogram, $e_s=-2s+2$, for the $2d$ 10-state Potts 
 models on various lattice sizes (re-drawn after Ref.~[3] from where 
 the notation for $s$ comes).} 
 \label{fig_potts2d10q_ln}
\end{figure}

It is assignment {\tt a0501\_06} to produce similar data for the $2d$ 
10-state Potts model and to re-weighted the multicanonical energy
histogram to the canonical distribution at $\beta=0.71$, which is close
to the pseudo-transition temperature. The multicanonical method allows 
then to estimate the interface tension of the transition by following 
the minimum to maximum ratio $R(L)$ of Eq.~(\ref{Itension}) over many 
orders of magnitude~\cite{BeNe92} as is shown in 
Fig.~\ref{fig_potts2d10q_ln}.

\subsection{Free Energy and Entropy Calculation}

\begin{figure}[tb] \vspace{5pc}
 \centerline{\hbox{ \psfig{figure=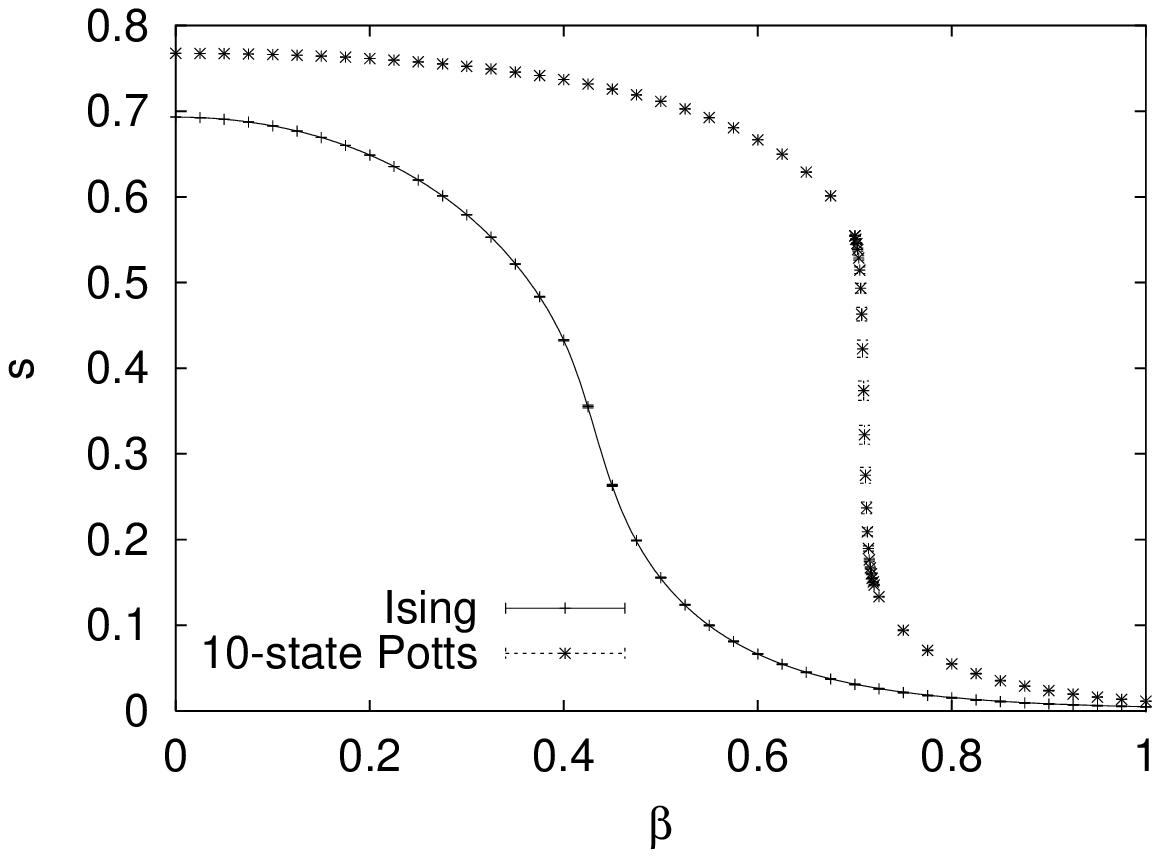,width=10cm} }}
 \caption{Entropies per spin, $s=S/N$, from multicanonical simulations 
 of the Ising and 10-state Potts models on an $20\times 20$ lattice 
 (assignments {\tt a0501\_03} and {\tt a0501\_05}). The full 
 line is the exact result of Ferdinand and Fischer for the 
 Ising model.} \label{fig_entropy}
\end{figure}

At $\beta=0$ the Potts partition function is $Z=q^N$. Therefore, 
multicanonical simulations allow for proper normalization of the 
partition function, if $\beta=0$ is included in the temperature 
range.
The properly normalized partition function allows to calculate 
the\index{Helmholtz free energy}\index{free energy} 
{\bf Helmholtz free energy}
\begin{equation} \label{free_energy} 
 F\ =\ - \beta^{-1}\, \ln ( Z )
\end{equation}
and the \index{entropy}{\bf entropy}
\begin{equation} \label{entropy} 
 S\ =\ {F - {E}\over T}\ =\ \beta\, (F - {E})
\end{equation}
of the canonical ensemble. Here ${E}$ is the expectation 
value of the internal energy and the last equal sign holds because of 
our choice of units for the temperature.
For the $2d$ Ising model as well as for the $2d$ 10-state Potts model, 
we show in Fig.~\ref{fig_entropy} multicanonical estimates of the 
entropy density per site
\begin{equation} \label{entropy_den} 
 s\ =\ S/N\ .
\end{equation}

For the $2d$ Ising model one may also compare directly with the number
of states. Up to medium sized lattices this integer can be calculated 
to all digits by analytical methods~\cite{Be95}. However, MC results
are only sensitive to the first few (not more than six) digits and, 
therefore, one finds no real advantages over using other physical 
quantities.

\subsection{Time series analysis}

Typically, one prefers in continuous systems time series data over 
keeping histograms, because one avoids then discretization errors
\cite{Berg}. Even in discrete systems time series data are of 
importance, as one often wants to measure more physical quantities 
than just the energy. Then RAM storage limitations may require to 
use a time series instead of histograms. To illustrate this point,
we use the Potts \index{magnetization}magnetization.

In assignments {\tt a0501\_08} and {\tt a0501\_09} we create the same 
statistics on $20\times 20$ lattices as before, including time series 
measurements for the energy and for the Potts magnetization. For 
energy based observables the analysis of the histogram and the time 
series data give consistent results. 

\begin{figure}[tb] \vspace{5pc}
 \centerline{\hbox{ \psfig{figure=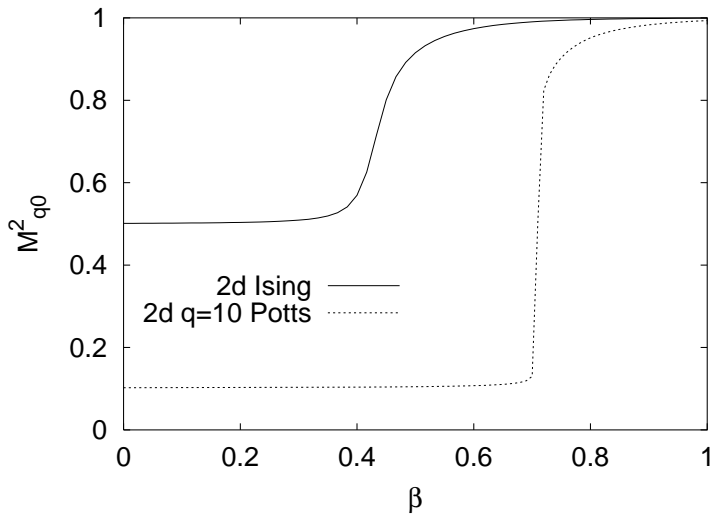,width=10cm} }}
 \caption{The Potts magnetization per lattice site squared for the 
 $q=2$ and $q=10$ Potts models on a $20\times 20$ lattice (assignments
{\tt a0501\_08} and {\tt a0501\_09}). } \label{fig_Potts_M2}
\end{figure}

For zero magnetic field, $H=0$, the expectation value of the Potts 
magnetization on a finite lattice is is simply
\begin{equation} \label{Mq_exp}
M_{q0}\ =\ \langle\, \delta_{q_i,q_0}\, \rangle\ =\ {1\over q}\ ,
\end{equation}
independently of the temperature. For the multicanonical simulation 
it is quite obvious that even at low temperatures each Potts state 
is visited with probability $1/q$. In contrast to this, the expectation 
value of the magnetization squared 
\begin{equation} \label{M2q}
M_{q0}^2\ =\ q\, \left\langle \left( {1\over N}\,\sum_{i=1}^N 
\delta_{q_i,q_0} \right)^2 \right\rangle
\end{equation}
is a non-trivial quantity. At $\beta=0$ its value is 
$M^2_{q0}=q\,(1/q)^2=1/q$, whereas it approaches $1$ for $N\to\infty$,
$\beta\to\infty$. For $q=2$ and $q=10$ Fig.~\ref{fig_Potts_M2} shows 
our numerical results and we see that the crossover of $M^2_{q0}$ 
from $1/q$ to $1$ happens in the neighborhood of the critical 
temperature. A FSS analysis would reveal that a singularity 
develops at $\beta_c$, which is in the derivative of $M^2_{q0}$ 
for the {second order phase transitions} ($q\le 4$) and in $M^2_{q0}$ 
itself for the {first order transitions} ($q\ge 5$).

\section*{Acknowledgments}
I like to thank Professor Louis Chen and the IMS staff for their kind 
hospitality. While visiting the IMS I greatly benefitted from 
discussions with Professors Wolfhard Janke, David Landau, Robert 
Swendsen and Jian-Sheng Wang.

\printindex

\end{document}